%% file: DistrSeqCtrlAttacks_TASE19.tex
\documentclass[journal]{IEEEtran}
\usepackage{graphicx}
\graphicspath{{Figures/}}

\usepackage{fancyvrb} 

\usepackage{amssymb}
\newcommand*{\QEDE}{\hfill\ensuremath{\square}}%

\newtheorem{remark}{Remark}
\newtheorem{example}{Example}

\newtheorem{property}{Property}

\usepackage{tabularx}

\usepackage{multirow}

\usepackage[export]{adjustbox}

\usepackage{todonotes}

\ifCLASSINFOpdf
\else
\fi
\begin{document}
\VerbatimFootnotes
\title{Security Analysis for Distributed IoT-Based Industrial Automation}%
%
%

\author{Vuk~Lesi,~\IEEEmembership{Student Member,~IEEE,}
        Zivana~Jakovljevic,~\IEEEmembership{Member,~IEEE,}
        and~Miroslav~Pajic,~\IEEEmembership{Senior Member,~IEEE}
\thanks{V. Lesi and M. Pajic are with the Department
of Electrical and Computer Engineering, Duke University, Durham, NC, 27708 USA (email:
vuk.lesi@duke.edu; miroslav.pajic@duke.edu.}
\thanks{Z. Jakovljevic is with University of Belgrade, Faculty of Mechanical
Engineering, Department for Production Engineering, 11000 Belgrade, Serbia (e-mail: zjakovljevic@mas.bg.ac.rs).}%
\thanks{Manuscript composed December 15, 2019.}
}

\maketitle

\begin{abstract}
With ever-expanding computation and communication capabilities of modern embedded platforms, Internet of Things (IoT) technologies enable development of Reconfigurable Manufacturing Systems---a new generation of highly modularized industrial equipment suitable for highly-customized manufacturing. Sequential control in these systems is largely based on discrete events, while their formal execution semantics is specified as Control Interpreted Petri Nets (CIPN). Despite industry-wide use of programming languages based on the CIPN formalism, formal verification of such control applications in the presence of adversarial activity is not supported.
Consequently, in this paper we focus on security-aware modeling and verification challenges for CIPN-based sequential control applications. Specifically, we show how CIPN models of networked industrial IoT controllers can be transformed into Time Petri Net (TPN)-based models, and composed with plant and security-aware channel models in order to enable system-level verification of safety properties in the presence of network-based attacks. Additionally, we introduce realistic channel-specific attack models that capture adversarial behavior using nondeterminism. Moreover, we show how verification results can be utilized to introduce security patches and motivate design of attack detectors that improve overall system resiliency, and allow satisfaction of critical safety properties. Finally, we evaluate our framework on an industrial case study.\end{abstract}

\renewcommand\abstractname{Note to Practitioners}
\begin{abstract}
Our main goal is to provide formal security guarantees for distributed sequential controllers. Specifically, we target smart automation controllers geared towards Industrial IoT applications, that are typically programmed in C/C++, and are running applications originally designed in e.g., GRAFCET (\mbox{IEC 60848})/SFC (\mbox{IEC 61131-3}) automation programming languages. Since existing tools for design of distributed automation do not support system-level verification of relevant safety properties, 
we show how security-aware transceiver and communication models can be developed and composed with distributed controller models. Then, we show how existing tools for verification of Time Petri Nets can be used to verify relevant properties including safety and liveness of the distributed automation system in the presence of network-based attacks. To provide an end-to-end analysis as well as security patching, results of our analysis can be used to deploy suitable firmware updates during the stage when executable code for target 
controllers (e.g., in C/C++) is generated based on GRAFCET/SFC control models. We also show that security guarantees can be improved as the relevant safety/liveness properties can be verified after corresponding security patches are deployed. 
Finally, we~show applicability of our methodology on a realistic distributed pneumatic manipulator. 

\end{abstract}


\renewcommand\abstractname{Primary and Secondary Keywords}
\begin{abstract}
Primary Topics: Sequential control systems, Secure distributed automation, Industrial Internet of Things;
Secondary Topics: Petri nets, Non-deterministic analysis
\end{abstract}

%
\IEEEpeerreviewmaketitle

\VerbatimFootnotes

\input{Intro}

\input{ProblemStatement}
\input{Modeling}

\input{AttackModel}
\input{Verification}
\input{CaseStudies}

\input{Conclusion}



%



\section*{Acknowledgment}
This work is sponsored in part by the ONR under agreements N00014-17-1-2012 and N00014-17-1-2504, as well as the NSF CNS-1652544 grant. It was also partially supported by Serbian Ministry of Education, Science and Technological Development, research grants TR35004 and TR35020.

\ifCLASSOPTIONcaptionsoff
  \newpage
\fi



%
%
%

\bibliographystyle{IEEEtran}
{
    \bibliography{DistrSeqCtrlAttacks_TASE19}
}

%








\end{document}

%% file: Intro.tex

\section{Introduction}
\label{sec:intro}

Advanced  capabilities of smart Internet of Things (IoT) devices 
have lead to their  widespread adoption in industrial automation system,
rapidly advancing reconfigurable manufacturing~\cite{jakovljevic_icamm17}; the rise of the fourth industrial revolution, known as Industry 4.0~\cite{I4.0}, introduces the new era of highly-customized (rather than highly-serialized) manufacturing~\cite{Hoda}. In this vision, manufacturing resources are highly modularized, providing the necessary flexibility to adapt to dynamical market demands. Efficient structural and functional changes are supported by Reconfigurable Manufacturing Systems (RMS) that can be configured \emph{ad-hoc} with little or zero downtime~\cite{Koren2018121}.

The foundation of RMS are modules controlled by smart Industrial IoT (IIoT)-enabled controllers. IIoT endpoints (sometimes referred to as \emph{industrial assets}) are heterogeneous by definition---they represent multi-vendor components whose deployment environment dynamically changes depending on the process needs and current configuration of RMS. Furthermore, a plethora of different communication technologies (wired and wireless) and protocols are employed~\cite{Boyes20181}. Seamless reconfiguration, integration and reliable functioning of RMS requires that components are highly autonomous. Specifically, they must be capable of seamlessly communicating with each other using compatible protocols (integrability), exchanging both low-level control-related and high-level process-bound information (interoperability) and to interact with each other in different ways to enable operation in a plethora of configurations (composability)~\cite{IIRA}.

Reconfigurability is naturally supported by distributed control architectures; conventionally centralized controllers are responsible of all aspects of control---from low-level event signaling, to high-level coordination. Their complexity hinders reconfigurability both from the hardware perspective (i.e., requiring component re-wiring), and the software aspect (i.e., having to ensure the control software is aware of and functions correctly under the new hardware configuration). Thus, the new generation of smart manufacturing resources must exploit not only functionally-required components (such as sensors and actuators) but also intrinsic computation and communication capabilities of IIoT-enabled controllers in order to enable a higher level of automation and autonomy. Control distribution enables decoupling of fine-grained details about \emph{how} control over the specific physical resource is performed, from the resources coordination problem which only needs to worry about \emph{what} the manufacturing resources are capable~of~performing.

\begin{figure}[t]
	\centering
	\includegraphics[width=0.5\textwidth]{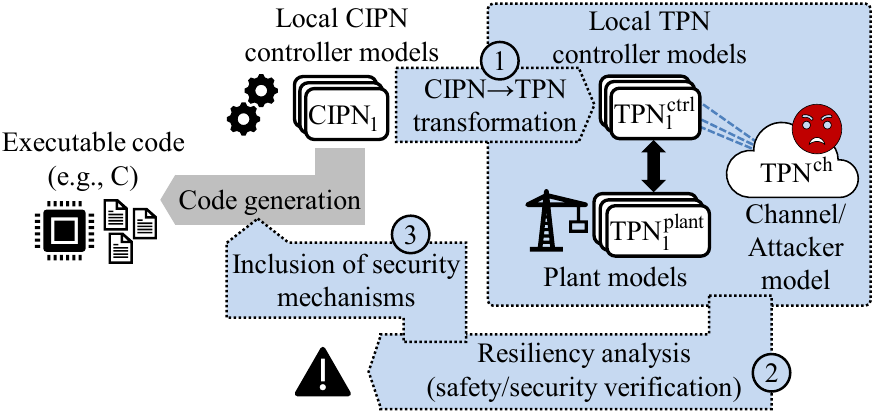}
	\caption{Our methodology for resilient IIoT-based distributed automation---in Phase 1, the composition of existing distributed control models, which are used to generate executable code for IIoT controllers, with channel and plant models is used to formally verify properties of interest in Phase 2. Finally, in Phase 3, the results of the security analysis are used to enhance system resiliency, by adding suitable  security mechanisms during code generation.}
	\label{fig:methodology}
\end{figure}

However, the networked nature of the new generation of distributed automation systems makes them susceptible to network-based attacks~\cite{wang_arxiv19}, similar to security vulnerabilities reported in other cyber-physical systems domains (e.g.,~\cite{pajic_csm17}). For example, an adversary may inject false events~\cite{802.15.4auth}, delay or deny network access to legitimate controllers~\cite{wirelessattack}, or manipulate control commands~\cite{ackattack2} sent over unsecure communication channels. On the one hand, providing security guarantees is critical in distributed sequential control systems where progress is directly impacted by communication unavailability. Yet, despite devastating effects such attacks could have on operation of distributed industrial automation systems, existing approaches to securing such systems are somewhat ad-hoc; commonly, the benefits of included security mechanisms for control performance (i.e., Quality-of-Control---QoC) are unclear and hard to evaluate if no formal system analysis can be performed.
Consequently, to enable building of secure and correct-by-design RMS, in this work we introduce efficient techniques for systematic security analysis of distributed control applications deployed on IIoT-enabled local controllers (LCs). We also show how results of the security analysis can be used to improve automation performance and safety guarantees in the presence of attacks, by adding suitable security mechanisms that address the detected vulnerabilities. This results in the overall framework, shown  in Fig.~\ref{fig:methodology}, for formal safety analysis and patching of distributed sequential automation systems under adversarial influences.

%

Coordination between components in a large fraction of IoT systems is based on discrete events. While a plethora of formal modeling frameworks is employed under the umbrella of IoT (e.g.,~\cite{iotsat,iotz3,iotautomata}), industrial automation systems are commonly based on GRAFCET (IEC~60848)/SFC (IEC~61131-3) control designs, and consequently on the underlying formal semantics of Control Interpreted Petri Nets (CIPN). Therefore, we focus on formal security analysis of IIoT-enabled controllers that are described using CIPNs; such controllers may be  developed directly, or automatically derived using methods for distribution of existing centralized sequential automation designs (e.g.,~\cite{jakovljevic_tcst19}), which allow for deployment of legacy control applications over IIoT-enabled smart controllers.

While inherent determinism of CIPNs is not a limitation when the formalism is used to specify controllers' behaviors, it prevents the use of CIPNs to model malicious actions~\cite{wang_arxiv19}. On the other hand, the sister formalism of Time Petri Nets (TPN) supports nondeterminism, which makes it a great candidate for security-aware modeling.
Thus, for the first phase of our security analysis, we introduce methods for
automatic transformation of domain-specific CIPN-based controller specifications (i.e., designs) into TPN-compliant representations. These TPN models enable closed-loop system modeling and analysis, by composing them with corresponding non-deterministic plant and security-aware communication channel models; we show how such security-aware models can be developed with the desired level of abstraction that allows us to capture impacts of attacks on automation performance. While our framework generally supports any communication channel implementation, we focus on the IEEE~802.15.4-based implementation featured in our evaluation setup.

In Phase 2, we employ open verification tools (e.g.,~\cite{romeo}) to perform system-wide verification of safety and QoC-relevant properties in the presence of attacks, based on the aforementioned security-aware closed-loop system model; it is important to highlight that we make no assumptions about the attacker's choice among all possible malicious actions nor the times when they (i.e., attack actions) may occur.
By enabling security analysis within the same family of formalisms (i.e., using a formalism that is closely related to the formalism used to design controllers),   we provide convenient domain-specific interpretation of analysis results. This allows us to exploit verification results in Phase 3 to orchestrate security patches in code generation which is performed based on original CIPN-based models.
Finally, we show the applicability of our methodology on a real-world industrial case study---a security analysis of an IIoT-enabled manipulator system. 

{
Specifically, the contributions of this work are as follows:
\begin{itemize}
  \item Security-aware framework for verification of system-level properties for distributed discrete-event controllers (based on CIPNs) in the presence of network-based attacks;
  \item TPN-based non-deterministic modeling of network-based attacks on distributed controller communication, with emphasis on capturing impacts on automation performance;
  \item Extension of the control software development cycle from security-aware analysis to firmware patching, in order to ensure correct operation in the presence of attacks;
  \item Full-stack proof-of-concept case study based on industry-grade components demonstrating applicability of the developed secure automation framework.
\end{itemize}
}

%

This paper is organized as follows. Sec.~\ref{sec:relatedWork} gives an overview of relevant related work, while Sec.~\ref{sec:motivation} provides the problem definition with emphasis on distributed IIoT-based automation. Sec.~\ref{sec:modeling} introduces TPN-based security-aware modelings and Sec.~\ref{subsec:channelAndCtrlChannelInteraction} derives the security-aware communication model. Specification and verification of relevant formal properties is presented in Sec.~\ref{sec:verification}, as well as the loop closure from verification to code generation to include security patches and improve system resiliency. Industrial case studies are discussed in Sec.~\ref{sec:evaluation}, before concluding remarks (Sec.~\ref{sec:conclusion}).


\section{Related Work}
\label{sec:relatedWork}
In~\cite{faruque_cpsattacks}, a model-based approach for simulating attacks on CPS is presented, but no formal verification is supported and experimental results are obtained based on specific attack implementations. In~\cite{faruque_survivability}, additional formal security assessment of industrial CPS controllers is performed, but analysis remains constrained to high-level vulnerabilities at the level of functional models. On the other hand, a comprehensive formal security analysis of wireless IoT communications under a specific attack model is presented in~\cite{wirelessIOTsecurity}, but no relations to implications for Quality-of-Control of the underlying physical process are considered.

Security analysis techniques for other IoT domains have also been recently proposed. In~\cite{soteria}, smart home IoT applications are formally surveyed for anomalous behaviors. However, formal adversarial models and implications of security vulnerabilities on  system operation, such as QoC and safety in the presence of attacks, are not considered. Similarly, \cite{celikiotguard} introduces a dynamic policy-based enforcement system for securing against unauthorized and unwanted control scenarios, but focuses only on architectures and platforms for consumer IoT applications mostly in smart home automation. In~\cite{iotsat}, an SMT-based framework for IoT security analysis is presented; yet, only abstract threat models are used, and the software architecture of IoT nodes is masked by behavioral modeling.

Note that Petri Nets (PNs) have been used for security-aware modeling and analysis.
For example, penetration analysis using attack trees was formalized through 
PNs (e.g.,~\cite{PNattackNets1,PNattackNets2}).
Coordinated cyber-physical attack modeling for smart grids was done in~\cite{PNgridAttacks}, but high-level attack scenarios were modeled, and the structure of the system components was coarsely abstracted. Modeling of risks and vulnerabilities towards avoidance and discovery for Unix-like software was performed (e.g.,~\cite{PNunixAttacks}) but without specifics of the underlying software architecture. \cite{SPNattacksCPS,SPNattacksGRID}~adopt stochastic PN-based attack models for CPS threats, while in~\cite{lesi_iotdi19}, authors formulate a framework for formal reliability analysis of networked IIoT sequential control applications based on CIPNs. On the other hand, \cite{petrinets_faults}~deals with fault detection in systems modeled by PNs. However, fault/failure models are limited to stochastic behaviors that cannot accurately capture adversarial actions (as described in~\cite{wang_arxiv19}). While cooperation and communication protocols were modeled with PNs (e.g.,~\cite{PNprotocols1,PNprotocols2})
, to the best of our knowledge, nondeterminism in Petri nets was not exploited for adversarial modeling.

%% file: ProblemStatement.tex


\section{Motivating Example and Problem Description}
\label{sec:motivation}

We first introduce CIPNs and the mother formalism of PNs, before presenting an illustrative distributed control application, used as a running example in this work to highlight security vulnerabilities caused by automation (i.e.,~control)~distribution. 

\begin{figure}[!t]
	\centering
	\includegraphics[width=0.462\textwidth]{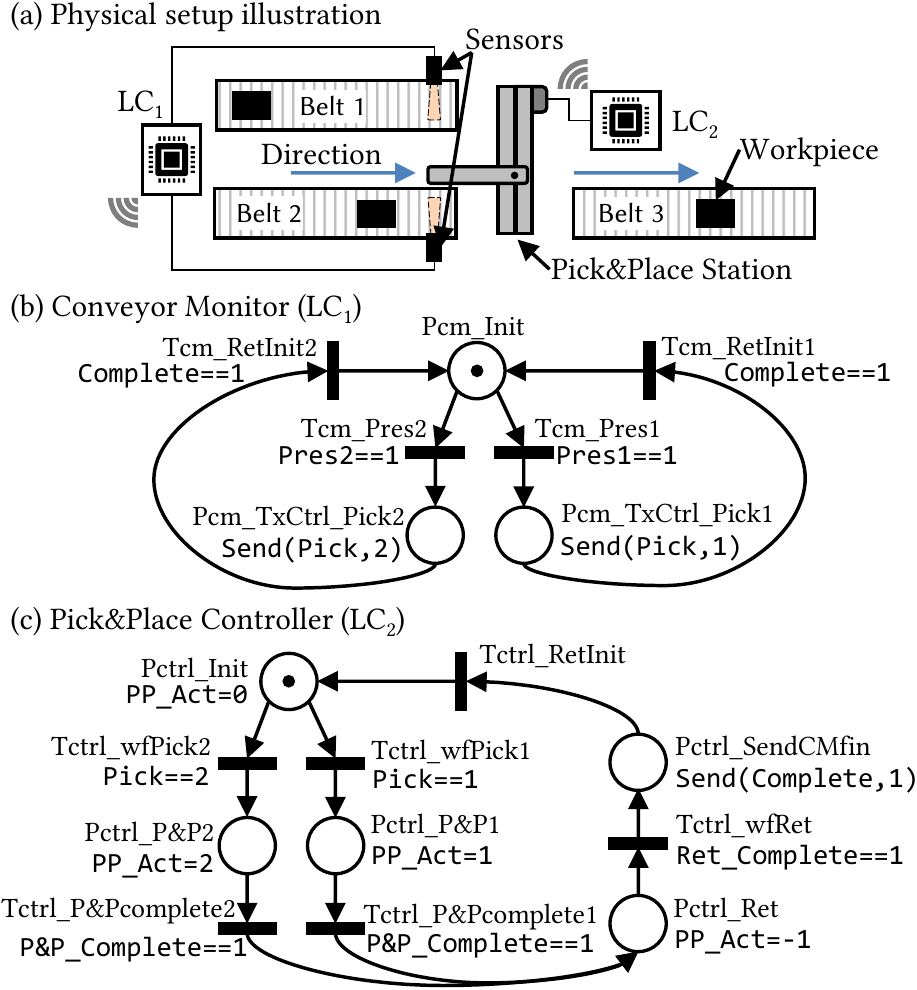}
	\caption{Distributed Automation Example: Simple CIPN-based distributed control model of (b) conveyor monitor~and (c)~pick~\&~place controller; physical setup is illustrated in~(a).}
	\label{fig:exampleCIPN}
\end{figure}

\subsubsection*{Petri Nets (PNs)}
A Petri net is a 5-tuple $\mathbf{PN}=(P,T,F,W,M_0)$, where $P=\{P_1,...,P_m\}$ is a set of places (represented by circles), $T=\{T_1,...,T_n\}$ is a set of transitions (represented by bars) such that $P\cup T \neq \emptyset$ and $P\cap T = \emptyset$, while $F\subseteq\{P \times T\}\cup \{T\times P\}$ is the set of arcs between places and transitions (no arc connects two places or two transitions). $W$ is the vector of arc cardinalities which determines how many tokens are removed/deposited over specific arcs upon firing of corresponding transitions. The $\mathbf{PN}$'s state  is defined by its \emph{marking}, i.e., distribution of tokens (captured by dots inside places); $M_0$ is the initial marking (i.e., the initial token distribution). Functionally, current PN marking determines the system's state, while transition firing (i.e., token flow) represents a state change. A formal description of the PN semantics can be found~in~\cite{Murata1989541}.

\subsubsection*{Distributed Automation with CIPNs}
CIPNs are a version of PNs where arc cardinality is fixed to $1$ and the initial marking $M_0$ may initialize only one place with a token. In CIPNs, transitions' firing can be conditioned by system inputs (i.e., sensors) in the form \verb!sensor==value!, while actuation commands can be associated with places in the form \verb!actuator=command!. For distributed automation, functionality of each local controller (LC) is captured by the corresponding CIPN.
For (event) information exchange between LCs, places of a CIPN may invoke the communication API exposed by the LC runtime environment; 
this is denoted as \verb!Send(signal,value)! for broadcast or \verb!Send({dest1,dest2,...},signal,value)! for uni/multicast transmissions. Dually, the receiving LC can condition its transitions with statements similar to conditioning on locally connected sensors (i.e., as \verb!signal==value!)~\cite{jakovljevic_tcst19}.

Formally, a CIPN is a 6-tuple $\mathbf{CIPN}=(P,T,F,C, A, M_0)$ where $P$, $T$, $F$, and $M_0$ are defined as for PNs; $C=\{C_1,...,C_n\}$ is a set of logical conditions 
enabling synchronization of the controller with sensors by guarding corresponding transitions in the $\mathbf{CIPN}$ model; 
$A=\{A_1,...,A_m\}$ is a set of actions 
on actuator outputs that are allocated to places; formal CIPN semantics is available in~\cite{David20101}. By its definition, CIPN semantics is deterministic (does~not support nondeterminism due to CIPNs use to only model controllers), which needs to be ensured during model~design~\cite{david1994petri}. 

Distributed CIPN-based controller models are obtained directly by design, or by distributing existing  (i.e., centralized) controllers 
(e.g.,~as done in~\cite{jakovljevic_tcst19}). 
Fig.~\ref{fig:exampleCIPN} shows a simple control application, built with two wireless nodes, that we will use as a 
running example in this work. 

\begin{example}
\label{ex:motivation}
Consider a simple application from~Fig.~\ref{fig:exampleCIPN}.
Control over the physical system (Fig.~\ref{fig:exampleCIPN}(a)) is performed by the conveyor monitor ($LC_1$) and the pick~\&~place station controller ($LC_2$). Two sensors (e.g.,~proximity, retro-reflective) locally connected to $LC_1$ overlook two parallel incoming conveyor belts; they sense if a workpiece is ready to be picked from either of the conveyors 
and placed on the third, outgoing conveyor.
The CIPN-based controllers 
for $LC_1$ and $LC_2$ are shown in Fig.~\ref{fig:exampleCIPN}(b-c).
Initially, $LC_1$ is in state \verb!Pcm_Init! where it is waiting for either of 
its sensors to indicate workpiece presence (i.e.,~transition \verb!Tcm_Pres1! / \verb!Tcm_Pres2! is conditioned by the sensing event \verb!Pres1==1! / \verb!Pres2==1!).\footnote{For model readability, we employ descriptive notation for~places, transitions, conditions, and actions; e.g., transition \verb!Tctrl_wfRet! in controller $LC_2$ waits for the return cycle to finish, while place \verb+Pcm_TxCtrl_Pick1+ on the conveyor monitor sends signal \verb+Pick==1+ to the pick~\&~place controller.} Upon detection of a workpiece (i.e.,~when one of 
\verb!Pres1==1! / \verb!Pres2==1! is satisfied, and thus transition \verb!Tcm_Pres1! / \verb!Tcm_Pres2! is enabled), $LC_1$ sends a message to $LC_2$ 
(via API call \verb!Send(Pick,1)! / \verb!Send(Pick,2)! in place \verb!Pcm_TxCtrl_Pick1! / \verb!..._Pick2!); 
the message 
indicates which conveyor has a workpiece ready to be picked. $LC_1$ then waits for completion of the pick~\&~place operation.

Concurrently, $LC_2$'s initial state is \verb!Pctrl_Init! where the pick~\&~place station is commanded to halt (by \verb!PP_Act=0!). Once the signal \verb!Pick! is received from $LC_1$, based on its value the token in $LC_2$ model transitions over \verb!Tctrl_wfPick1! / \verb!Tctrl_wfPick2! into place \verb!Pctrl_P&P1! / \verb!Pctrl_P&P2! where the corresponding actuation command is given to the pick~\&~place station (i.e.,~\verb!PP_Act=1! / \verb!PP_Act=2!). After completion of the pick~\&~place operation, condition \verb!P&P_Complete==1! is satisfied, allowing the $LC_2$ to transition over \verb!Tctrl_P&Pcomplete! / \verb!Tctrl_P&Pcomplete! to \verb!Pctrl_Ret! where it commands the pick~\&~place station to return to home position (by \verb!PP_Act=-1!). $LC_2$ waits for completion of the pick~\&~place station return stroke (when 
\verb!Ret_Complete==1! evaluates to true); after it transitions back into \verb!Pctrl_SendCMfin!, where it signals 
$LC_1$ that the workcycle is complete. $LC_2$ transitions over \verb!Tctrl_RetInit! back into the initial place \verb!Pctrl_Init!.
Finally, conveyor monitor $LC_1$ can also return to its initial state (formally, the token is deposited back into \verb!Pbm_Init! over \verb!Tbm_RetInit1! / \verb!Tbm_RetInit2!), as the condition \verb!Complete==1! is satisfied.\QEDE
\end{example}

The CIPN-based control models from Example~\ref{ex:motivation} assume ideal communication (i.e., packet delivery), without unpredictable 
channel behaviors. For instance, consider an adversary with network access that mounts an \emph{impersonation} (i.e., spoofing or masquerade~\cite{wirelessattack}) attack 
when $LC_2$ is waiting for a message from $LC_1$ 
that a workpiece should be picked up. By sending 
the corresponding message (e.g., by signaling \verb!Send(Pick,1)!), the attack will result in the pick~\&~place station  being commanded pickup by $LC_2$; hence, it may collide with upcoming workpieces, potentially incurring mechanical damage, or just waste a workcycle. Similar 
 holds for \emph{message modification}~\cite{wirelessattack} (i.e., signal replacement~\cite{DESsupervisoryControl}) attacks,
 when the right conveyor belt contains a workpiece ready to be picked up (i.e., \verb!Pres_R==1!), 
 but the attacker intercepts the corresponding message and maliciously signals \verb!Send(Pick,2)!. 
Also, if an adversary \emph{delays} or \emph{blocks} some transmissions or acknowledgements (ACKs)~between LCs (i.e., launching a Denial-of-Service (DoS) attack~\cite{wirelessattack}) 
the system may experience excessive downtime.

These examples illustrate that distributing control and automation functionalities may introduce security vulnerabilities as the system operation can be easily affected by an attacker with network~access.
Hence, in this work, we focus on security aspects of 
IIoT-enabled distributed automation systems; our goal is to provide techniques to model and analyze system behaviors in the presence of  \emph{network-based attacks}, while enabling the use of analysis results 
to modify (i.e., update)) the system to achieve attack-resilient operation.

\subsection{Overview of our Approach}
We start from a functional description 
of $N$ LCs expressed by $\mathbf{CIPN}_i$, $i=1,...,N$. We consider an attacker with full access to the network with {$M$ communication channels}. 
The attacker is not able to compromise LCs, but has full knowledge of the state of each~LC. 
We introduce a design-time methodology illustrated in Fig.~\ref{fig:methodology} that starts with automatic transformation of CIPN-based control models to Timed Petri Net~(TPN)-based models; such models allow for explicit capturing of (i)~the communication semantics, (ii)~platform-based effects using timed transitions to model real non-zero execution and message propagation times, and most importantly (iii)~non-deterministic behaviors necessary to model adversarial actions. 

We show how the remaining closed-loop system components (i.e., the plant and communication channel in the presence of attacks) can be modeled within the TPN formalism. Furthermore, we  demonstrate how composition of these models enables system-wide analysis of control performance in the presence of attacks. Finally, we show how design-time formal verification results can be used during code generation for smart IIoT-based controllers, which facilitates adaptation  of LCs' firmware in order to address exposed security~concerns. 

\begin{remark}[Petri nets vs. Automata/Finite-State Machines]
We employ Petri Net-based modeling formalisms since CIPNs are the main formalism used to capture existing (including distributed) automation  systems. 
For example, GRAFCET~(IEC~60848) and SFC (IEC 61131-3) languages for programming
industrial control systems originate from Petri Nets, with their behavioral equivalence discussed in~\cite{David20101,GRAFCETtoTPN,SFCtoTPN}.
However, 
the proposed methodology, including network and attack modeling, can be directly extended to other discrete-event IoT systems whose behavior can be captured with automata/finite-state machines due to the fact that formal mappings between semantics of Petri nets and automata have been defined (e.g.,~as in~\cite{PNtransformation1,PNtransformation2}).\QEDE
\end{remark}

%% file: Modeling.tex

\section{TPN-Based  Automation Modeling} 
\label{sec:modeling}

TPNs extend PNs by introducing timed transitions; in a TPN, every transition is characterised by an interval $(\underline{t}_f,\overline{t}_f),[\underline{t}_f,\overline{t}_f),(\underline{t}_f,\overline{t}_f],$ or $[\underline{t}_f,\overline{t}_f]$, where $\underline{t}_f$ and $\overline{t}_f$ are the lower/upper bound on the transition firing times, 
which may be \emph{zero} or \emph{infinity}---time interval next to \emph{immediate} transitions (i.e., with zero firing time) is not specified.
Also, 
the firing times are defined relative to the moment of transition enabling, without any assumptions 
on their distribution. 
This enables modeling of timed properties of real-time control software~\cite{TPNforRT1,TPNforRT2,TPNforRT3}.
In addition, by supporting non-determinism, TPNs facilitate attack modeling 
that cannot be accurately done with deterministic or stochastic models. 


Therefore, we transform formal distributed control specifications expressed by $\mathbf{CIPN}_i$, $i=1,...,N$, into TPN-compatible models $\mathbf{TPN}_i^{ctrl}$, $i=1,...,N$. We then compose these models with plant models $\mathbf{TPN}_i^{plant}$, $i=1,...,N$, and security-aware communication channel models $\mathbf{TPN}_j^{channel}$, $j=1,...,M$,\footnote{This captures the general case where time or frequency multiplexing may be used to provide multiple communication channels over the same medium.} which enables us to reason about system-level safety and security properties under the modeled adversarial influences. Since both CIPNs and TPNs originate from PNs, the translation from $\mathbf{CIPN}_i$, $i=1,...,N$ controller models to $\mathbf{TPN}_i^{ctrl}$, $i=1,...,N$ is direct for all places and transitions except where platform-implemented API is called,  i.e., 
\begin{enumerate}
  \item Places handling actuation, and transitions handling sensing by issuing I/O API calls 
  (\verb!actuator=value! and \verb!sensor==value!, respectively),
  \item Places handling transmissions over a shared channel, and transitions handling receiving of communication signals 
  using API calls (\verb!Send(destination,signal)! and \verb!signal==value!, respectively),
  \item Places calling other platform-dependent API, such as request for execution delays.
\end{enumerate}
These CIPN constructs, which directly rely on the underlying platform used to implement the controller , 
must be explicitly modeled as nets that capture: 1)~interaction between LCs $\mathbf{TPN}_i^{ctrl}$ and the plant $\mathbf{TPN}_i^{plant}$, 2)~interaction between LCs $\mathbf{TPN}_i^{ctrl}$ and communication channel(s) $\mathbf{TPN}_j^{channel}$, and 3)~runtime environment 
changes based on issued commands (e.g., variable updates, execution delays).

Thus, we introduce methods for automatic extraction of TPN-based controller models from existing CIPN-based models that may be used to generate control code. We capture interaction between the automation system and plant, as well as the platforms' runtime environment in Section~\ref{subsec:plantAndControllerInteraction}, 
and security-aware modeling of communication channels and their interfaces with controllers in Section~\ref{subsec:channelAndCtrlChannelInteraction}. These methods enable developing a complete closed-loop system model that can then be used to reason about system resiliency to~attacks.

\subsection{Modeling Plants and Controller-Plant Interaction}
\label{subsec:plantAndControllerInteraction}
Nominal behavior of the controlled physical system is typically known at control design time. Thus, since the formalism of CIPNs is universally adopted for automation design, we assume that a PN-based (i.e., $\mathbf{CIPN}$ or $\mathbf{TPN}$) plant model is available.\footnote{On the other hand, if an automata or other discrete-event system representation of the plant is available, existing tools and methodologies can be used to translate such models into a PN-based representation (e.g.,~\cite{PNtransformation1,PNtransformation2}).}
On the running example, we describe development of such TPN-based plant model, along with a TPN-compliant controller-plant interface implemented through marking-dependent \emph{guard functions}. 



%
\begin{figure}[t]
	\centering
    \includegraphics[width=0.46\textwidth]{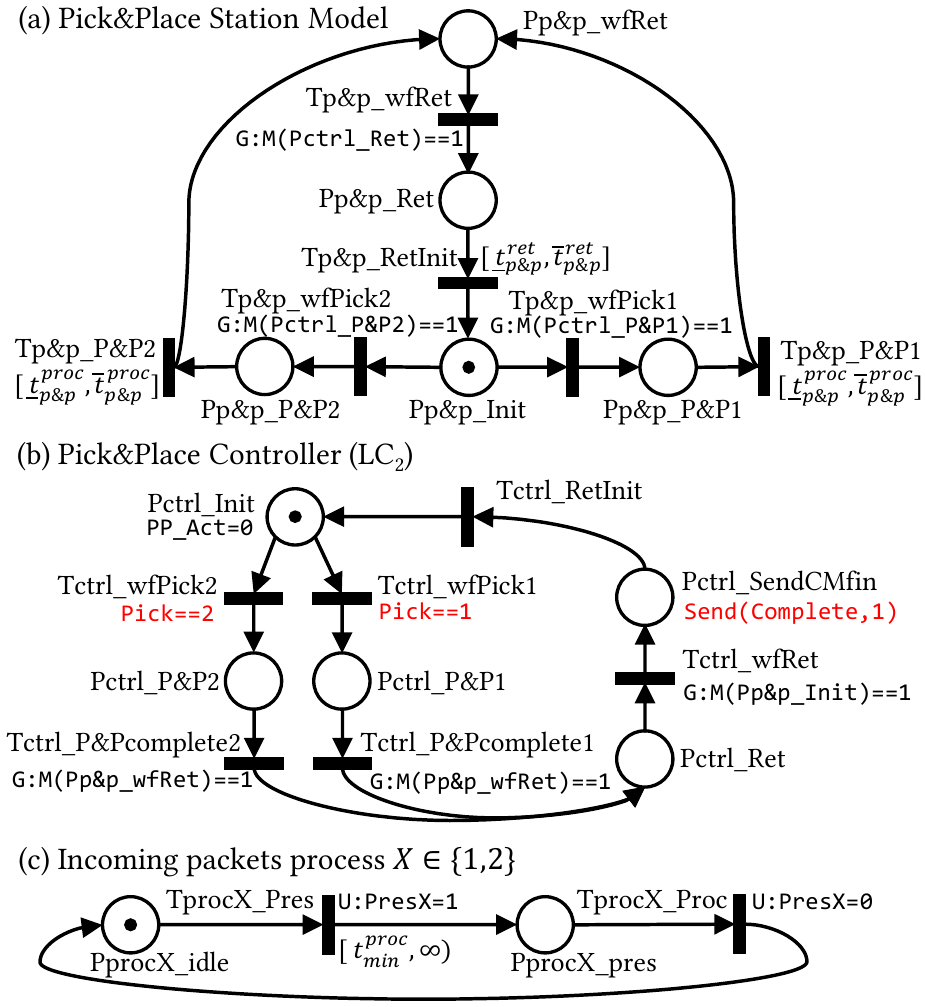}
	\caption{Plant and plant-controller interaction modeling: (a) $\mathbf{TPN}$ model of the pick~\&~place station; 
	(b) extended model of $LC_2$ from~Fig.~\ref{fig:exampleCIPN}(c) with TPN-compatible sensing/actuation -- the model is not a $\mathbf{TPN}$ as it still relies on the communication API (in {\color{red}red}) for interaction with other LCs; (c) model of incoming workpieces with a lower bound on the workpiece interarrival.}
	\label{fig:ctrlPlantInteraction}
\end{figure}

Fig.~\ref{fig:ctrlPlantInteraction}(a) shows a $\mathbf{TPN}$ model of the pick~\&~place~station from Fig.~\ref{fig:exampleCIPN}; Fig.~\ref{fig:ctrlPlantInteraction}(c) models  the incoming workpieces arrivals, with a lower bound on the interarrival times.
 Place \verb!Pp&p_Init! represents the station's initial state. Token flow from this place is conditioned by the corresponding commands of $LC_2$ \verb!PP_Act==1! and \verb!PP_Act==2! from the controller model in Fig.~\ref{fig:exampleCIPN}(c). In the formalism of TPN, \emph{marking-dependent guard functions} can be used to restrict state changes (i.e., token flow) in the plant model, i.e., a marking-dependent function (denoted with \verb!M(!$\cdot$\verb!)!) assesses the state of the controller model (i.e., token distribution) and returns the current number of tokens inside the argument place. Hence, guard function \verb!G:M(Pctrl_P&P1)==1! (or \verb!G:M(Pctrl_P&P2)==1!  for the other conveyer belt) is associated with transition \verb!Tp&p_wfPick1! (or \verb!Tp&p_wfPick2!).
Once the pick~\&~place process is triggered (i.e., $LC_2$ initiates it from a specific conveyor belt) by the $LC_2$'s model advancing its token to \verb!Pctrl_P&P1! (or \verb!Pctrl_P&P2!), the station's token transitions to place \verb!Pp&p_P&P1! (or \verb!Pp&p_P&P2!) if conveyor belt 1 (or 2) has a workpiece waiting to be processed.
Note that in order to capture realistic executions, the actual times to complete the pick~\&~place and return processes are not deterministic; this is natively supported by TPN's timed transitions; transitions \verb!Tp&p_P&P1! / \verb!Tp&p_P&P2! have firing times from the interval $[\underline{t}_{p\&p}^{proc},\overline{t}_{p\&p}^{proc}]$, as shown in Fig.~\ref{fig:ctrlPlantInteraction}(a).\footnote{These intervals may be obtained experimentally under nominal conditions at runtime, or based on design constraints 
at design-time. Also, different durations of the picking process from different conveyors are supported but set equal in the above model/figure to simplify our presentation.} 

Now, the station dwells in place \verb!Pp&p_wfRet! waiting for the signal from $LC_2$ to return to the home position, i.e., guard function \verb!G:M(Pctrl_Ret)==1! conditions transition \verb!Tp&p_wfRet! on the corresponding controller's state (i.e., where the return action is issued). Once return is commanded, the pick~\&~place station takes non-deterministic time from 
$[\underline{t}_{p\&p}^{ret},\overline{t}_{p\&p}^{ret}]$ to return (i.e., transition over \verb!Tp&p_RetInit!). After this transition, 
the station model is back in the initial state, waiting to process the next workpiece. 

The actuation part of the plant-controller interface is managed by guard functions assessing the controller's marking; therefore, explicit actuation input updates (e.g., \verb!PP_Act=1!) in the CIPN places are omitted in the transformation to the TPN model as TPN places do not feature any attributes. This interface can also be achieved alternatively, by utilizing \emph{update functions} which are triggered on the firing of controller's transitions and can update markings or variables. The choice of the transformation semantics from CIPN to TPN can therefore be adjusted to the specific platform implementation.




Dually to the actuation part of the interaction, 
sensing is modeled by introducing plant-marking-dependent guard functions on controller's transitions. Specifically, transitions conditioned by sensor values in the form \verb!sensor==value! in the CIPN controller model are replaced with immediate transitions guarded by a Boolean function evaluating to \emph{true} if the plant model marking corresponds to the plant state where \verb!sensor==value! is satisfied, and to \emph{false} otherwise.

For instance, once $LC_2$ commands return of the pick~\&~place station, (i.e., controller model from Fig.~\ref{fig:exampleCIPN}(c) has the token in \verb!Pctrl_Ret!), it is blocked on the transition \verb!Tctrl_wfRet! which is guarded by the condition \verb!Ret_Complete==1!. This transition in the CIPN model is transformed into a transition in the TPN model in Fig.~\ref{fig:ctrlPlantInteraction}(b) that is guarded by a function dependent on the marking of the pick~\&~place station model from Fig.~\ref{fig:ctrlPlantInteraction}(a)~(i.e., controller waits for the station to reach home position). Guard function \verb!G:M(Pp&p_Init)==1! returns \emph{true} when the token in the plant model transitions from \verb!Pp&p_Ret! to \verb!Pp&p_Init! over the timed transition \verb!Tp&p_RetInit! (here, \verb!M(!$\cdot$\verb!)! denotes a function that returns the current number of tokens inside the argument place) -- hence, controller $LC_2$ can transition over \verb!Tctrl_wfRet!. Therefore, this guard function is used for the transition \verb!Tctrl_wfRet! in the TPN model of $LC_2$. This models the controller side of the controller-plant interaction, i.e., sensor sampling. More complex conditions based on multiple sensors are implemented by forming an arbitrary plant marking-dependent Boolean guard function.

Fig.~\ref{fig:ctrlPlantInteraction}(b) shows a controller model of 
the described controller-plant interface. 
The model, obtained from the CIPN-based model in~Fig.~\ref{fig:exampleCIPN}(c), is intermediary, and not fully TPN-compliant; the CIPN-based communication semantics (i.e., signal transmissions via \verb!Send(destination,signal)!, and receptions through \verb!signal==value! denoted in red in Fig.~\ref{fig:ctrlPlantInteraction}(b)) is still present in the model. {However, to allow for verification of properties for systems where networking is not a concern, this communication semantics can be easily adapted to TPNs by applying the same \emph{guard/update} functions as described; this results in a model architecture from Fig.~\ref{fig:modeling}(a).

Additionally, note that the conveyor monitor model from Fig.~\ref{fig:exampleCIPN}(b) can be directly transformed into a TPN, with guards \verb!Pres1==1! and \verb!Pres2==1! conditioning progress based on external inputs (i.e., the process) triggered by the process model shown in Fig.~\ref{fig:ctrlPlantInteraction}(c). This process model abstracts away the nature of the process of incoming workpieces on the conveyors with a minimum inter-arrival time (i.e., time in the interval $[t_{min}^{proc},\infty]$). 

\subsubsection{Controller Runtime Environment Modeling}
\label{sec:runtimemodel}

Another challenge for the automatic mapping of CIPN-based control models into TPN-compliant models is mapping of places issuing system calls from the runtime environment (e.g., execution delays, requests for timer interrupts, setting counter events) or updating local controller state (e.g., manipulating global variables). Requested execution delays can easily be modeled as timed transitions with the exact firing times (i.e., where the lower and upper firing time bound are the same); in general, however, event timings with different semantics are available depending on the control implementation---i.e., GRAFCET or SFC. In~\cite{GRAFCETtoTPN,SFCtoTPN}, authors provide detailed translational semantics between CIPNs and TPNs in these cases by introducing \emph{event sequencers} as certain conditions exist where transitions can be taken while time to some events generated internally in places has still not elapsed.
\begin{remark}[Modeling more Complex Execution Environments]
While we consider single-threaded automation examples (as most sequential control implementations are), existing techniques for modeling parallel systems can be applied given the expressiveness of TPNs. For example, for multithreaded applications where task preemption is allowed, the operating system scheduler can be modeled as a separate component, even in case of multi-processor platforms~\cite{TPNscheduling}. 
%
\QEDE
\end{remark}

\subsection{CIPN and TPN Controller Equivalence}
  An execution path in CIPNs can be defined as a sequence of markings, where a change in the marking occurs due to firing of a transition. Recall that places are associated with actions; hence, each marking is associated with a set of actions, while transitions are associated with guards---firing of each transition is thus conditioned by a set of conditions.\footnote{We employ the standard assumption that all inputs are re-evaluated after firing of every transition (e.g., as done in~\cite{SFCtoTPN}).} Therefore, an execution path is a sequence $M_0,\mathcal{T}_1,M_1,\mathcal{T}_2,...$, where $\mathcal{T}_{i}$ is the transition taking the net from marking $M_i$ to $M_{i+1}$. In the TPN model, a path is characterized by a similar sequence with the addition of transition timing.\footnote{Strictly speaking, two types of time intervals characterize each transition: static intervals (i.e., design-time bounds) when they may fire, and dynamic (i.e., runtime) intervals when they can fire at any given instant, conditioned by all other enabled transitions. However, for purposes of showing marking-based equivalence with CIPNs, time can be abstracted away.} In our case, it is sufficient to maintain the CIPN \emph{controller} execution paths in the TPN model, as our objective is operational equivalence of the source (centralized), and the target (distributed) control models.

{
CIPN controller specifications are fully deterministic by design, and have only \emph{immediate} transitions.\footnote{Immediate transitions become fireable immediately after enabling, i.e., without a time delay.}
Thus, the corresponding target TPN-based controller models obtained by directly constructing the same model in the TPN formalism without additional constructs (i.e., as previously described in Sec.~\ref{subsec:plantAndControllerInteraction}), 
do not introduce behavior which is not covered by the source CIPN-based models. Consequently, execution paths of the composition of the TPN models match with that of the CIPN, from the input-output (i.e., sensing-actuation) perspective. In other words, no execution path is added when we transform the CIPN-based controller into the TPN-based representation~\cite{jakovljevic_tcst19}. Intuitively, the TPN models obtained by direct mapping from CIPN (i.e., place-by-place, and transition-by-transition), are still \emph{fully deterministic} (isolated from the intrinsically non-deterministic plant and channel); i.e., their behavior is identical to their CIPN counterparts, and same behavioral assumptions (e.g., 1-boundedness) hold~\cite{David20101}.
}


%
\begin{figure}[t]
	\centering
    \includegraphics[width=0.45\textwidth]{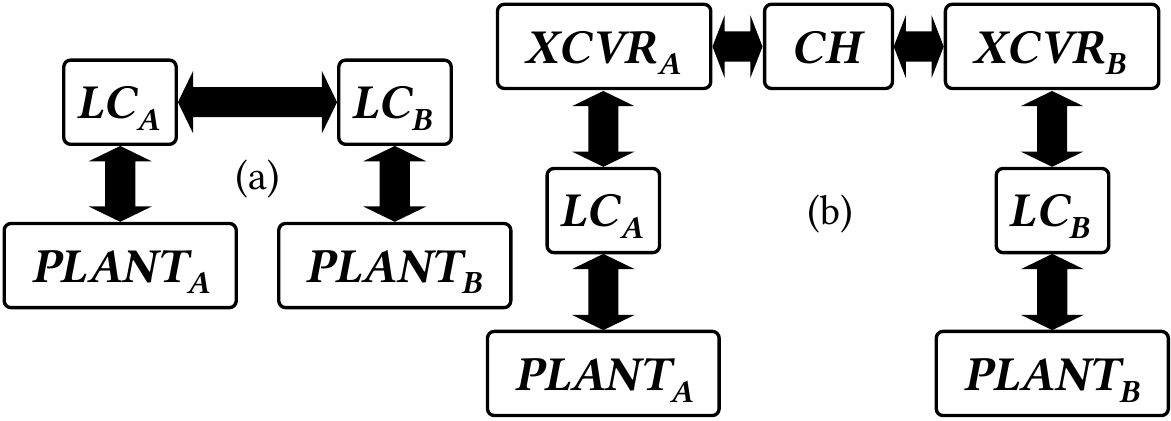}
	\caption{Model architecture: (a) Model captures local controllers $LC_i$, plants $PLANT_i$ and their interactions, (b) Model also captures the employed communication transceivers $XCVR_i$, and the underlying communication channel $CH_i$. Note that local controller models  $LC_i$ in schemes (a) and (b) are not the same; i.e., in (b), controller places/transitions invoking communication APIs are made compatible with the transceiver model.}
	\label{fig:modeling}
\end{figure}

%% file: AttackModel.tex

%
\begin{figure*}[t]
	\centering
	\includegraphics[width=0.98\textwidth]{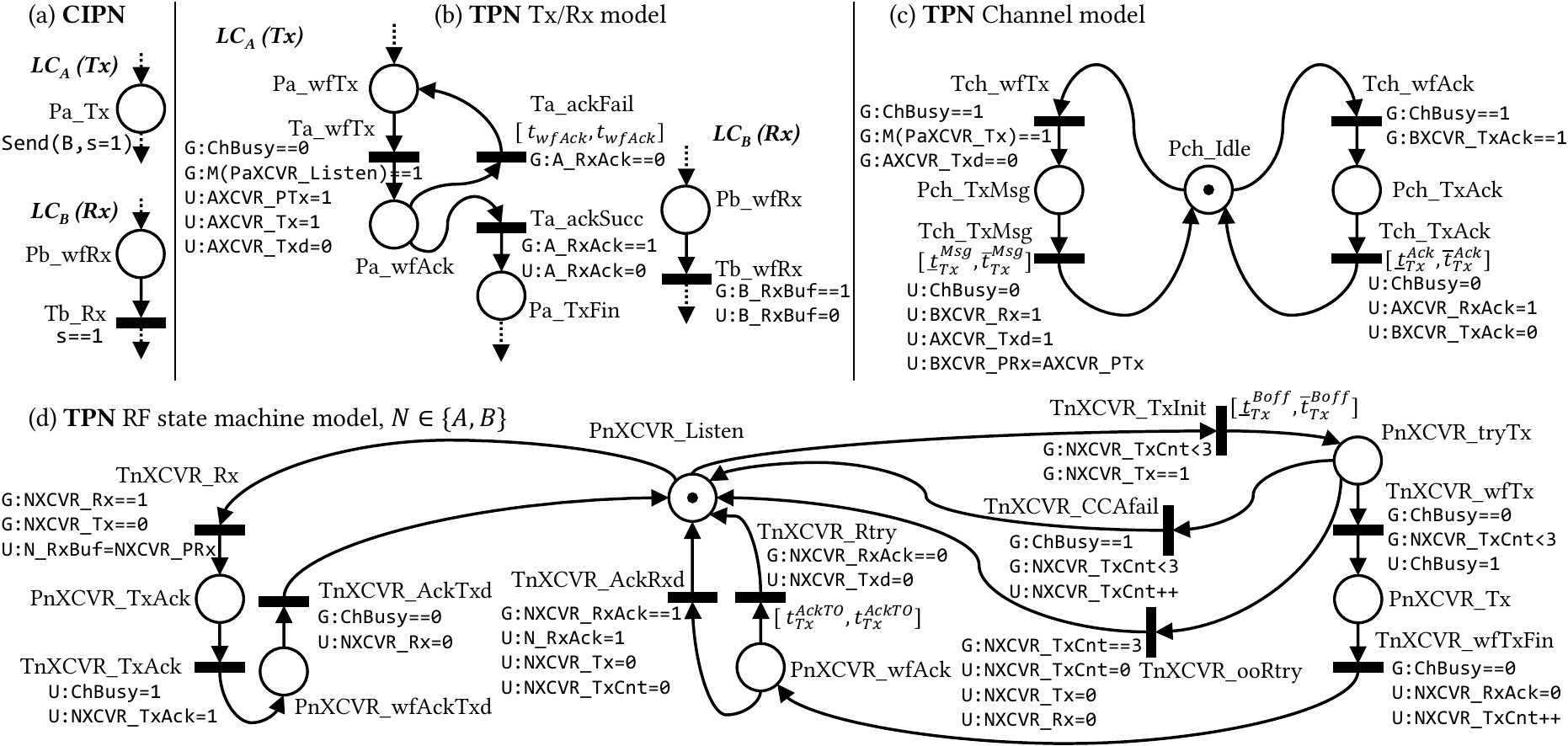}
	\caption{Transformation between CIPN-based and TPN-compatible communication models; (a) a Tx/Rx place/transition pair in the CIPN formalism; (b) the same Tx/Rx place/transition pair modeled with as a TPN adjusted to the half-duplex, acknowledge-required unicast CSMA-CA-based channel, whose model is shown in (c); (d) model of the employed radio transceiver (i.e., the governing RF state machine TPN model). Note that each Tx/Rx net pair in (a) (from the model in Fig.~\ref{fig:modeling}(a)) is extended into a corresponding pair in (b), while only a single model from (c) and (d) are added to obtain the model from Fig.~\ref{fig:modeling}(b).}
	\label{fig:CIPNtoTPN}
\end{figure*}

\section{Security-Aware Modeling of the Channel  and Controller-Channel Interaction}
\label{subsec:channelAndCtrlChannelInteraction}

We now introduce a security-aware channel model, including a TPN-compliant controller-channel interface that enables model composition. Hence, we address modeling challenges to enable the transition from the security-agnostic model structure from Fig.~\ref{fig:modeling}(a), to the security-aware model composition shown in Fig.~\ref{fig:modeling}(b). We start by defining the attack model.

\subsection{Attack Model}
\label{subsec:attackModel}
We assume 
a powerful network-based adversary that:
\begin{enumerate}
  \item Has the full knowledge of the distributed system, 
  including the CIPN models, generated code and analysis framework, as well as  the current state of all LCs (and their transceivers).
  \item Has network access and full communication protocol compliance, i.e., is able to transmit unsigned messages as any of the LCs, or intercept messages or 
ACKs exchanged by~LCs,
  \item Is able to precisely time actions and align transmissions with legitimate network traffic, e.g., to interfere with legitimate messages by transmitting the carrier signal or a protocol-compliant message.
\end{enumerate}

Thus, the adversary may mount the following attacks:
\begin{enumerate}
  \item \emph{Interception or delaying of legitimate packets (DoS):} With these attacks, adversarial transmissions occupy the channel (a)~blocking transmissions 
  from legitimate LCs to prevent or delay their access to the network, 
  or (b)~blocking ACKs on legitimate LC's transmissions to cause unnecessary retransmissions and slow down progression of the targeted transmitter~\cite{wirelessattack}.

  \item \emph{ACK spoofing:} The attacker may impersonate an ACK expected by a legitimate transmitter; e.g., following by interception of the transmission, the attacker may spoof the ACK misleading the legitimate transmitter into believing that the transmitted signal was received by the intended receiver~\cite{802.15.4auth}, both for regular and `heartbeat'/sync messages~\cite{ackattack1}.

  \item \emph{Impersonation/Masquerade:} 
  The adversary may transmit false event signals on behalf of a legitimate LC (i.e., impersonating another controller), with the goal to inject false commands~\cite{802.15.4auth} or sensor measurements; such attacks could e.g., allow the targeted receiver to resume execution while it is blocked waiting for 
  an event signal, before the event 
   it is sent by the legitimate LC.

  \item \emph{Signal replacing/Message modification:} The adversary may modify content of a legitimate message 
  to deliver false event information. While logically the same, the attack procedure differs from intercepting a legitimate transmission followed~by a masquerading attack~\cite{ackattack2}, and thus is modeled~differently.\footnote{This type of attack is technically more challenging to perform compared to other attacks, especially over a wireless medium.}

  \item \emph{Replay attack:} The attack characterizes an adversary that records events signaled by the LCs and replays the sequence of events on behalf of one or more LCs; thus, maliciously emulating activity of LCs whose operation (s)he is interfering with~\cite{ackattack2}.
\end{enumerate}

The above attack set, considered in this work, covers all reported attacks that, from the standpoint of low-level signaling of events, could have direct impact on Quality-of-Control (QoC) of the underlying physical process~\cite{wang_arxiv19,wirelessControl}. 
Other attacks, such as attacks on {network routing policies}, are focused on higher-level information flows and are thus harder to directly relate to automation QoC~\cite{ackattack2}. Consequently, our goal is to model attacks by capturing their influence~on~the sequential control system and the resulting QoC, rather than the employed attack vector for any specific attacks; i.e., the attack model should be agnostic to the actual attack implementation.

\subsection{TPN-Based Modeling of Attack Impact}

Recall that CIPN models rely on platform-provided communication APIs for passing events between LCs; e.g., as in Fig.~\ref{fig:CIPNtoTPN}(a),
\verb!Send(destination,signal=value)! command within a place sends the updated \verb!value! of \verb!signal! to the \verb!destination! LC, while condition \verb!signal==value! on a transition within the model blocks execution until the signal corresponding to the desired value is received over the network. 
To 
enable formal analysis of the attack impact on QoC of distributed automation, it is necessary to develop a TPN-compliant model of the interface (i.e., transceiver) between the controller and security-aware channel model; such model can be then composed with the TPN-based models described in Section~\ref{subsec:plantAndControllerInteraction}, 
resulting in Fig.~\ref{fig:modeling}(b) architecture.

{
Such security-aware formal model has to capture:
(1)~application-level (i.e., controller side) communication stack behavior, directly affected by
(2)~the channel-side (i.e., communication medium) attack model, and
(3)~the controller-channel interface.
Specifically, application-level (i.e., control-related) communication stack behavior, such as delays or blocking on communication peripheral resources, is of interest for security analysis, as this presents the main reflection of the communication-level attacks onto the control functionality. Therefore, when translating the CIPN communication model from Fig.~\ref{fig:CIPNtoTPN}(a) into a TPN-compliant model, it is necessary to capture application software states that directly affect progress of the control functionality, conditioned by data dependencies resolved via communication. Such models can be obtained from the actual application firmware running on the embedded LCs (i.e., source code). For example, when IEEE~\mbox{802.15.4} protocol is used, as in the case study presented in Sec~\ref{sec:evaluation}, the state-machine/TPN representation can be directly extracted from the radio driver (as done in Fig.~\ref{fig:CIPNtoTPN}(b)).
On the other hand, if more complex communication stack is considered (i.e., implementing higher communication layers), exiting state-machine extraction techniques (e.g.,~\cite{soteria}) can be employed.

Second, the channel model has to explicitly capture the channel states essential for supporting the attack models presented in Section~\ref{subsec:attackModel}---channel states that are not observable (or alterable) need not be modeled (e.g., bit-level signaling, or carrier-level modulation). Finally, a TPN-compliant interface between the controller and security-aware channel models is needed to allow for their formal composition, 
enabling system-level analysis of adversarial influence on the entire system. Therein, specific data link layer (OSI model layer 2) features are crucial for understanding retransmissions and ACK mechanics which, as we will show, affects design of attack detectors. Therefore, while controller models should capture application-level communication semantics, it is also necessary to include protocol-level details within the transceiver (XCVR) models, which act as the interface between the controllers and the medium (as shown in Fig.~\ref{fig:modeling}(b)). XCVR specifics are commonly available for the specific employed radio communication chip as RF circuitry control is usually state-machine based (e.g.,~referred to as the \emph{internal RF state machine}~\cite{mrf} in the case of radios used in our implementation).

On the other hand, explicit security-aware channel modeling is \emph{medium-, protocol-, and attack-dependent}. Fig.~\ref{fig:CIPNtoTPN}(c) and Fig.~\ref{fig:attackModels} show a security-aware model of a \emph{half-duplex, acknowledge-required unicast CSMA-CA-based communication channel} with respect to the previously defined attack model. While other medium/protocol variants can be easily modeled due to the expressiveness of TPNs, we consider this specific model as it applies to our physical setup described later in Section~\ref{sec:evaluation}. In the remaining of this section, we describe the transformation from the CIPN-based LC communication model to a TPN-compliant model assuming the aforementioned channel, while aiming to balance between the model expressiveness and capturing security-aware behavior required for analysis of QoC under attack.
}


\begin{table}[t]
  \centering
    \begin{tabular}{p{1.6cm}|p{5.4cm}|p{0.4cm}}
      Symbol & Description & SW acc.\\ \hline\hline
      \verb!ChBusy! & Indicator whether the channel is currently busy with a packet or ACK & YES \\ \hline
      \verb!N_RxBuf! & Local Rx buffer & YES \\ \hline
      \verb!N_RxAck! & Local flag indicating successful transmission, i.e., ACK reception & YES \\ \hline
      \verb!NXCVR_PTx! & Transceiver Tx payload buffer & YES \\ \hline
      \verb!NXCVR_PRx! & Transceiver Rx payload buffer & YES \\ \hline
      \verb!NXCVR_Tx! & Signal to XCVR initiating transmission & YES \\ \hline
      \verb!NXCVR_Txd! & Signal to XCVR indicating transmission & NO \\ \hline
      \verb!NXCVR_Rx! & Signal from XCVR indicating reception & YES \\ \hline
      \verb!NXCVR_TxAck! & XCVR signal initiating ACK transmission & NO \\ \hline
      \verb!NXCVR_RxAck! & XCVR signal indicating ACK reception & NO \\ \hline
      \verb!NXCVR_TxCnt! & XCVR retry counter & NO \\ \hline
      $\underline{t}_{Tx}^{Msg},\overline{t}_{Tx}^{Msg}$ & Message transmission time (bounds) & --- \\ \hline
      $\underline{t}_{Tx}^{Ack},\overline{t}_{Tx}^{Ack}$ & ACK transmission time (bounds) & --- \\ \hline
      $\underline{t}_{Tx}^{Boff},\overline{t}_{Tx}^{Boff}$ & Back-off time (bounds) & --- \\ \hline
      $t_{Tx}^{AckTO}$ & Data link layer ACK timeout & --- \\ \hline
      $t_{wfAck}$ & Application-level ACK timeout & --- \\ \hline
      $\underline{t}_{Ch}^{DoS},\overline{t}_{Ch}^{DoS}$ & Contention time due to DoS (bounds) & --- \\ \hline\hline
    \end{tabular}
  \caption{Symbols used in Fig.~\ref{fig:CIPNtoTPN} and Fig.~\ref{fig:attackModels}; third column (where applicable) indicates accessibility to application software (or only to the transceiver's internal RF state machine).}\label{tab:CIPNtoTPNsymbols}
\end{table}

\subsection{Security-Aware Modeling of the Channel and Controller-Channel Interaction}

\begin{figure*}[t]
\vspace{-10pt}
	\centering
	\includegraphics[width=0.98\textwidth]{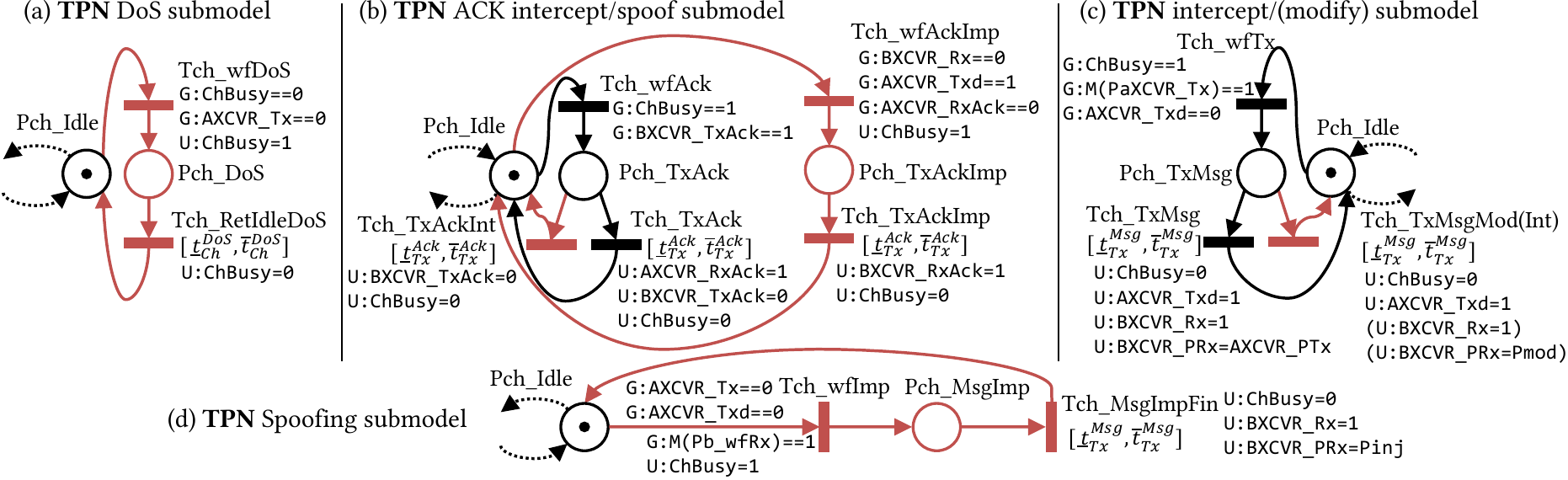}
	\caption{Additional places, transitions, and arcs required to obtain a security-aware channel model for different attack types: (a) DoS, (b) ACK intercept/spoof, (c) message modification, and (d) masquerade. All attack-related components of the model are depicted in red color, while nominal components are shown partially for completeness in black (where relevant).}
	\label{fig:attackModels}
\end{figure*}

Fig.~\ref{fig:CIPNtoTPN}(b) shows the TPN transmitter/receiver models that replace the platform-independent CIPN transmitter/receiver model in Fig.~\ref{fig:CIPNtoTPN}(a). Fig.~\ref{fig:CIPNtoTPN}(c) shows the nominal channel model (i.e., without adversarial influences), while Fig.~\ref{fig:CIPNtoTPN}(d) shows the transceiver (XCVR) model. Notice that both $LC_A$ and $LC_B$ have identical transceivers; thus, $N\in\{A,B\}$ in place/transition names. Table~\ref{tab:CIPNtoTPNsymbols} enumerates symbols (local flags, variables, and transition timing parameters) used in the models in Fig.~\ref{fig:CIPNtoTPN},~\ref{fig:attackModels}. The internal RF state machine can be in the \emph{listening}, \emph{transmitting a packet}, \emph{waiting for acknowledgement}, or \emph{transmitting an acknowledgement} states. The transceiver employed in our case study (in Sec.~\ref{sec:evaluation}), performs up to three retransmissions before signaling a transmission failure to the application. On the application level, an unbounded number of retransmissions are performed in case the transceiver returns failure. The TPN model in Fig.~\ref{fig:CIPNtoTPN}(b-d) models this interaction.

%
%
In the remaining of this section, we show how  the attacks described in Section~\ref{subsec:channelAndCtrlChannelInteraction} can be modeled as TPNs. Specifically, we describe additional places, transition, and arcs to be added to the nominal channel model shown in Fig.~\ref{fig:CIPNtoTPN}(c) to capture the attacks. To enhance model readability, Fig.~\ref{fig:attackModels} depicts only additional places and transitions in red color required to model a specific attack, while the nominal places and transitions are depicted in black (all parts of the nominal model not relevant for the specific attack are omitted therein).

\paragraph{DoS attack submodel} Fig.~\ref{fig:attackModels}(a) shows the DoS attack submodel. When the channel is idle, the attacker may decide to occupy the channel to prevent legitimate transmissions. He/she may do so at any time (non-deterministic choice) when the channel is not busy, and keep the channel busy arbitrarily long. In the model, the channel is kept busy for some non-deterministic time in the range $[\underline{t}_{Ch}^{DoS},\overline{t}_{Ch}^{DoS}]$, after which it is released by the attacker.

\paragraph{ACK interception/spoofing submodel} Fig.~\ref{fig:attackModels}(b) shows the ACK intercept/spoof submodel. To model ACK interception, an additional transition is needed allowing the channel to return to idle state following ACK transmission, without the transmitter ($LC_A$) receiving the ACK sent by the receiver ($LC_B$); i.e., transition \verb!Tch_TxAckInt! is added as shown in Fig.~\ref{fig:attackModels}(b), and is \emph{not} associated with the update function \verb!U:BXCVR_RxAck=1!.
However, when ACK spoofing is considered, the attacker may transmit an ACK when the targeted receiver \emph{is not} in the process of acknowledging, while the targeted transmitter \emph{is} in the process of waiting an ACK.

Additionally, malicious ACK spoofing may be performed when the signal to which the ACK is intended to correspond is transmitted already, but the ACK has not yet been received by the sender (e.g., due to an intercepted ACK from the legitimate receiver). This is enabled with the additional net branch in Fig.~\ref{fig:attackModels}(b) starting with transition \verb!Tch_wfAckImp!. As a result of firing of this transition, the channel is declared busy and the spoofed ACK is assumed to take the same time as transmitting legitimate ACKs; thus,  the transition \verb!Tch_TxAckImp! has the same attributes as \verb!Tch_TxAck!, with the exception of the signal to the targeted transmitter signalling ACK transmission is done (i.e., update \verb!U:AXCVR_RxAck=1! is omitted).

\paragraph{Message intercept/modify submodel} Fig.~\ref{fig:attackModels}(c) shows the message intercept/modify submodel, where an additional transition \verb!Tch_TxMsgMod! (\verb!Tch_TxMsgInt!), {represented as one transition for conciseness}, is added. In the case of the modification attack, this transition in the model allows the attacker to deliver a signal different form the one originally transmitted (i.e., \verb!U:BXCVR_PRx=Pmod! where \verb!Pmod! is the payload modified by the attacker). In the case of packet interception, no update to the receiver's XCVR receiver buffer is made, and consequently the XCVR is not notified of a received packet (i.e., update functions are omitted and denoted as \verb!(.)! in Fig.~\ref{fig:attackModels}(c)).

\paragraph{Message impersonation submodel} Fig.~\ref{fig:attackModels}(d) presents the masquerade submodel. The additional transitions and places allow the attacker to make a non-deterministic choice to impersonate transmission of the expected transmitter whenever the channel is not busy, the targeted receiver is waiting for the corresponding signal, and the original transmitter is not in the process of sending this signal. Then, similarly to the nominal (legitimate) transmission model (shown in Fig.~\ref{fig:CIPNtoTPN}(c)), the transmission takes a non-deterministic time in the same range as legitimate transmissions. Note that the received payload on $LC_B$ is in this case the value \verb!Pinj! crafted by the attacker, rather than \verb!AXCVR_PTx!, normally transmitted by $LC_A$ in the adversary-free case.

\begin{remark}[Replay attacks]
Due to the introduced non-determinism, any specific sequence of attack actions are contained within the presented model (as long as the individual actions correspond to the attack model from Section~\ref{subsec:channelAndCtrlChannelInteraction}). Thus, replay attacks are covered by the presented model as they are only specific executions of the presented security-aware channel model. On the other hand, using a similar approach, finite memory replay attacks can be captured by a model that restricts inserted attack signals only to the previously transmitted messages, as done in~\cite{wang_arxiv19}.\QEDE
\end{remark}
\begin{remark}[Controller-plant VS. controller-channel interface modeling fidelity]
Control interface to the channel is modeled in far more detail than the interface to the plant, by abstracting away locally-connected actuator drives, relays, analog amplifiers, etc. The reason is that, in this work, we do not consider physical plant-level attacks. Hence, modeling the controller-plant interaction at a lower level of abstraction would unnecessarily increase model complexity. However, the presented techniques can be easily extended and the framework fully adapted to also cover physical attacks on the plant.\QEDE
\end{remark}

%% file: Verification.tex
\section{Resiliency Analysis and Security Patching}
\label{sec:verification}
A security-aware closed-loop system model obtained by composing the developed security-aware TPN models can be used to verify system-level safety and QoC properties in the presence of attacks. TPN analysis tools (e.g.,~\cite{romeo,tina}) allow for verification of formal properties specified as Linear Temporal Logic (LTL), Computational Tree Logic (CTL), or Timed CTL (TCTL) formulas~\cite{modelChecking}. In this work, we employ the tool Romeo~\cite{romeo} that enables verification of TCTL-based formal queries, such as  traditional safety (e.g., 1-boundedness~\cite{David20101}) and liveness properties (e.g., absence of deadlock).
In addition, as plant models are included, we can specify relevant domain-related plant-state-bound properties that are crucial for functional safety and QoC assessment. For our running example, the considered properties include: 

\begin{property}\label{prop:p1}
  A workpiece on conveyor~1~never triggers a pick-up from conveyor~2; this can be formally captured as:
 \verb!AG(not(M(Pp&p_P&P2)==1 and! \verb! activeConveyor==1))!.\footnote{Variable \verb!activeConveyor! is set when a workpiece presence is detected (i.e., on transitions \verb!Tcm_Pres1! or \verb!Tcm_Pres2! of the conveyor monitor, shown as CIPN in Fig.~\ref{fig:exampleCIPN}) and reset when conveyor monitor returns to initial state (i.e., over \verb!Tcm_RetInit!)}, where \verb!A! and \verb!G! are quantifiers signifying formula satisfaction along \emph{all} paths and \emph{always} (i.e., along all subsequent paths), respectively.
\end{property}

\begin{property}\label{prop:p2}
  A workpiece detected on any of the conveyors is eventually picked-up: \verb!(M(Pcm_TxCtrl_Pick1)==1 or! \verb!M(Pcm_TxCtrl_Pick2)==1)-->(M(Pp&p_wfRet)==1)!, where \verb!-->! denotes the {"leads to"} property; i.e., \verb!p-->q! means that for all executions, continuous satisfaction of property \verb!p! implies always eventual satisfaction of property \verb!q!, or formally \verb!AG(p => AF(q))!.
\end{property}

\begin{property}\label{prop:p3}
  The pick~\&~place station does not commence cycle (i.e., it is neither in \verb!Pp&p_P&P1! nor \verb!Pp&p_P&P2!), while the conveyor monitor is waiting for incoming workpieces (i.e., in the place \verb!Pcm_Init!). Formally, \verb!AG(M(Pcm_Init)+M(Pp&p_P&P1)+M(Pp&p_P&P2)<=1)!.
\end{property}

Using the Romeo tool, we verified that these as well as other QoC- and safety-critical properties are \emph{not} satisfied in the presence of attacks, as the attacker is capable of significantly altering the intended interaction between LCs, at arbitrary moments in time. For example, Property~\ref{prop:p1} is violated under \emph{message modification} attacks, Property~\ref{prop:p2} under possible infinite \emph{DoS}, while Property~\ref{prop:p3} fails under \emph{spoofing}.
The aforementioned properties are violated regardless of the values of timing parameters used in the model. Note that bounds on time-to-transmit and time-to-acknowledge can be obtained from experimental measurements, or directly from network specifications. Also, transceiver-related timings (e.g., back-off time during clear channel assessment) can be obtained from the employed transceivers' specifications. 

Regarding verification scalability---in the system model, one nominal pick~\&~place cycle,~with all attacks disabled, contains around $35$ transitions, which is on the order of the number of states in the model. Model complexity increases with the addition of non-deterministic attack choices, besides the time-induced non-determinism (in the plant model).\footnote{Romeo does not output statistics of the state space underlying the model.} Yet, in all cases, the tool takes less than $1~s$ to find an execution path violating the properties, on a workstation with an Intel~i7-8086K CPU ($4~GHz$ clock) and $64~GB$~memory.

\subsection{Addressing the Discovered Vulnerabilities}
As our previous analysis have shown, attack actions may significantly affect performance of distributed IoT-based industrial automation systems; to address them, it is necessary to add certain security mechanisms. In this section, we discuss how such security mechanisms affect  system models and verifiability of the relevant~properties.

\begin{figure}[!t]
	\centering
	\includegraphics[width=0.48\textwidth]{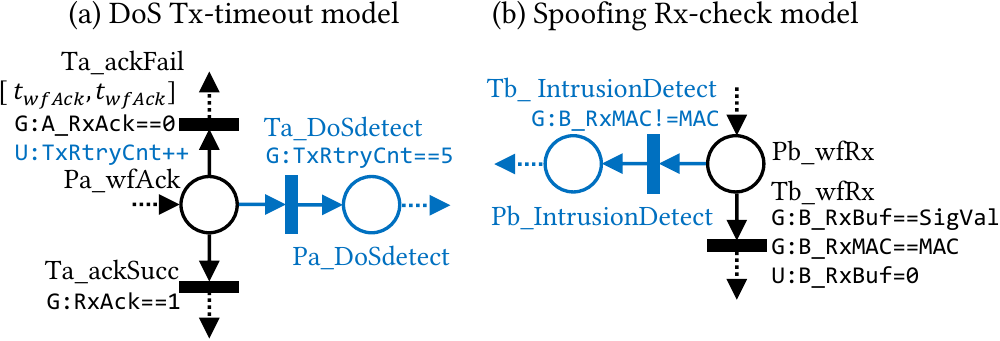}
	\caption{Model adaptation to addition of security services; (a) for DoS detection, and (b) against spoofing. Additional places and transitions are shown in blue color.}
	\label{fig:MACandDOSmodels}
\end{figure}
\begin{figure}[!t]
	\centering
	\includegraphics[width=0.44\textwidth]{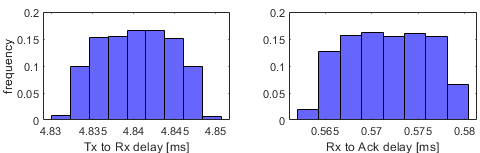}
	\caption{Tx-to-Rx and Rx-to-Ack times measured on our IEEE~802.15.4-enabled LC platform described in Section~\ref{sec:verification}.}
	\label{fig:histograms}
\end{figure}

\subsubsection{Detecting Denial-of-Service Attacks}
\label{subsubsec:DoSdetection}
Packet and acknowledgement (ACK) dropouts are common in wireless communication, and hence ACK and retransmission mechanisms are commonly used in such setups. For instance, in our experimental setup described in Sectio~\ref{sec:evaluation}, ACK request can be disabled in transceiver settings, in which case no retransmissions are attempted on the data link layer. For two isolated transceivers, this amounts to the one-way packet success rate of approximately $99~\%$ (see histograms in~Fig.~\ref{fig:histograms} that exclude unsuccessful transmissions). Thus, when ACK requests are enabled, up to three data-link layer retransmissions are performed,\footnote{The XCVR model in Fig.~\ref{fig:CIPNtoTPN}(d) is compatible with these specifications.} and we experimentally observed that no application-level retries are required beyond the three low-level protocol-provided retransmissions, in the case when a single industrial machine operates in isolation.

On the other hand, to increase network utilization, we emulated a number of additional machines communicating over the same wireless channel in physical vicinity (described in more detail in Section~\ref{sec:evaluation}); we experimentally observed the one-way packet success rate of approximately $98~\%$. Thus, two application-level retries were sufficient to enable reliable exchange of events, ensuring correct operation. Intuitively, protocol-provided retries are issued in short bursts while application-level retransmissions incur significant delay; the channel is more likely to be continuously busy for a short period of time (e.g., occupied by other legitimate transmissions). Yet, an adversary may repeatedly deny network access to legitimate controllers preventing the system from progressing. {Consequently, the modeled system does not satisfy Property~\ref{prop:p2}, despite application-level retransmissions, unless DoS attacks can be detected and system halted (or other precautionary actions taken), using e.g., a separate secure channel.}

From the operational perspective, every LC may implement a limited number of successive application-level retransmissions before declaring that it is under attack. For instance, if in our setup from Section~\ref{sec:evaluation}, we limited the number of retransmissions to five, amounting to a theoretical one-way packet success rate of \emph{eight nines} (if application-level retransmissions are assumed to be independent). To address this from the modeling perspective, we add an additional place where the transmitter's model transitions to, when application-level retries are exhausted (see Fig.~\ref{fig:MACandDOSmodels}(a)). Hence, we can verify that if infinite blocking of medium access is allowed, LCs may end up in the place \verb!Pa_DoSdetect!. Conversely, if DoS attacks are limited to four consecutive channel access denials, Property~\ref{prop:p2} is satisfied. Note that immediate emergency halt of the machinery may not be possible if a secure communication channel is not available or the DoS attacks cannot be isolated from the network (e.g., using bus guardians).

\subsubsection{Authenticating network flows}
\label{subsubsec:authentication}
Traditional cryptographic techniques for ensuring integrity of network flows rely on signing packets with Message Authentication Codes (MAC)~\cite{802.15.4auth}, and can be used to defend against spoofing attacks. In this setting, every transmission between LCs is signed by the transmitter using a secret key, and the signature is verified by the receiver; therefore, the attacker cannot tamper with the message payload, or else he/she will be detected.

From the modeling perspective, introducing authentication can be modeled as an additional condition on the receiving transitions (in the controller models) where the received payload is compared to desired values; i.e., the MAC portion of the payload is compared to a secret value that cannot be altered (in the case of modification) or generated (in the case of spoofing attacks) by the attacker. Specifically, the transition \verb!Tb_wfRx! in the $LC_B$ model in Fig.~\ref{fig:CIPNtoTPN}(b) would feature an additional guard function on the \verb!B_RxMAC! variable. Optionally, if the signature verification fails, a transition to a place modeling intrusion detection reaction can be added as shown in Fig.~\ref{fig:MACandDOSmodels}(b); this is left to the application designer as reacting to detected intrusions is highly application-specific.

Using the developed framework, we verified that if non-authenticated transmissions are not allowed (i.e., authentication implemented), Properties~\ref{prop:p1}~and~\ref{prop:p3} can be verified over our running example, under the condition that infinite denial of network access to LCs is not allowed, as previously discussed.

\subsubsection{Acknowledgement Spoofing}
Authenticating transmission does not affect ACKs as the data-link layer is responsible for ACK packets while MACs are added to the packet payload. In addition, non-encrypted sequence numbers, which are part of the packet frame, can be overheard by the attacker. Thus, valid ACKs can be generated on behalf of inactive (failed) LCs. Also, undelivered (i.e., intercepted) transmissions can be falsely acknowledged, even when authentication is used. This is a well-known shortcoming of data link layer ACKs~\cite{ackattack2,802.15.4sec}, and could be alleviated by application-level ACKs. Enforcing consensus over event-propagation in discrete event systems spans beyond the scope of this paper; yet, the presented modeling techniques can be utilized to model additional implemented protocols.

While this section introduced the general security-aware modeling aspects, with occasional focus on specific medium access techniques to avoid overly general discussions, in the following section we demonstrate the use of the presented framework on real-world industrial case studies.

%% file: CaseStudies.tex

\section{Case Studies: Industrial Manipulators}
\label{sec:evaluation}
We consider a full physical implementation of a reconfigurable industrial pneumatic manipulator with a variable number of modules/degrees~of~freedom (DOF) controlled in a distributed fashion; i.e., one local controller per module/DOF. We demonstrate effectiveness of our framework on multiple module configurations (i.e., 2-DOF, 3-DOF).

\begin{figure}[t]
	\centering
	\includegraphics[width=0.49\textwidth]{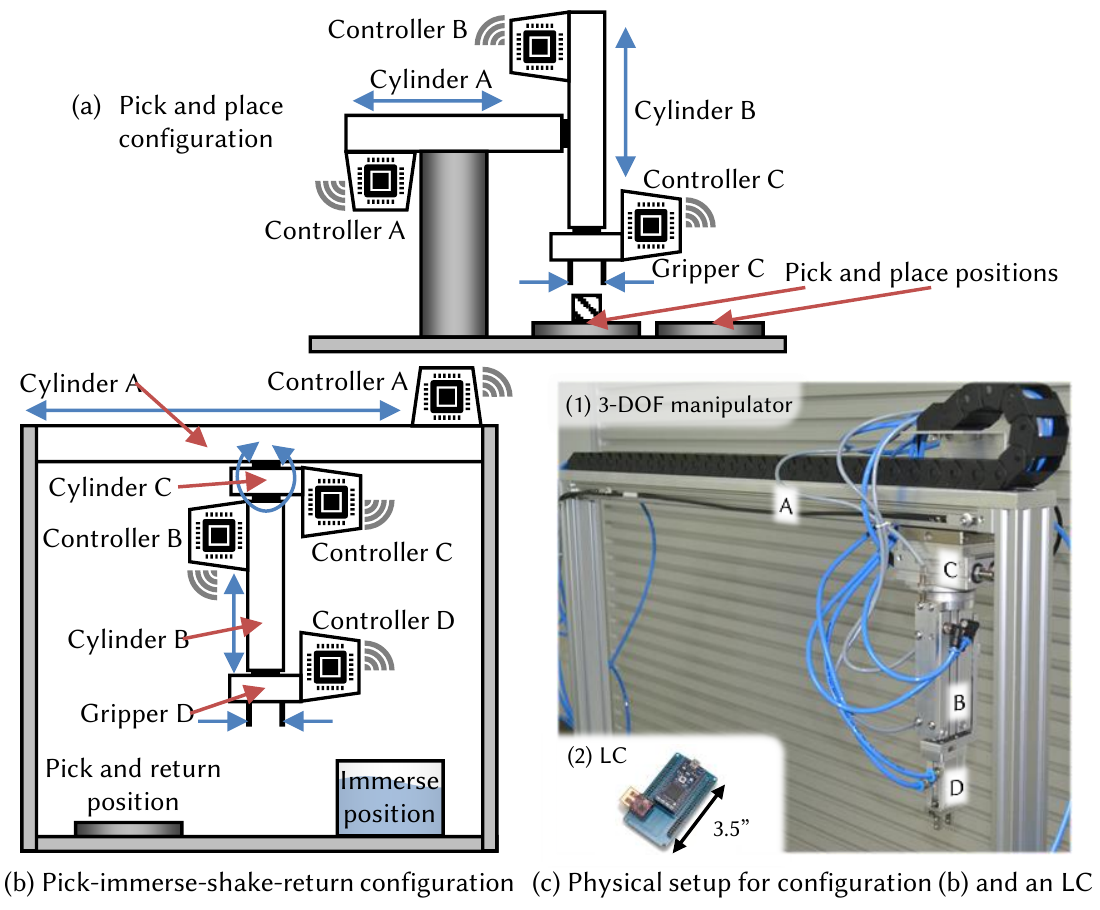}
	\caption{Pneumatic manipulator in multiple configurations: (a) 2-DOF~pick \&~place configuration; (b) 3-DOF pick-immerse-shake-return configuration; (c,1) upper portion of the physical setup of the configuration (b) shows cylinders; (c,2) low-cost ARM Cortex-M3-based networked controller; each physical component (cylinders and the gripper) are equipped with one LC.}
	\label{fig:caseStudies}
\end{figure}

\subsection{2-DOF Industrial Pneumatic Manipulator}
The pneumatic industrial manipulator in the 2-DOF configuration is depicted in Fig.~\ref{fig:caseStudies}(a); two double-acting cylinders (denoted $A$ and $B$) provide translational degrees of freedom, while the pneumatic gripper (denoted $C$) provides means of handling the workpiece. All actuation commands are issued by updating electrical signals \verb!xp!, $\verb!x!\in\{\verb!a!,\verb!b!,\verb!c!\}$ which activate monostable dual control pneumatic valves.\footnote{ A control valve is the interface between the controller and the pneumatic cylinder; it converts the actuation signal from the controller into mechanical movement that controls flow of pressured air towards pneumatic cylinders.} Notice that signals are denoted with \verb!x! while cylinders are denoted with $X$. Cylinders $A$ an $B$ are equipped with two proximity switches which allow position (i.e., fully retracted, fully extended) sensing. Signals corresponding to fully retracted (home) position are denoted \verb!x0!, while fully extended (end) position signals are denoted \verb!x1!. Additionally, the system contains a start switch whose corresponding signal is denoted by \verb!st!.

\begin{figure}[t]
	\centering
	\includegraphics[width=0.498\textwidth]{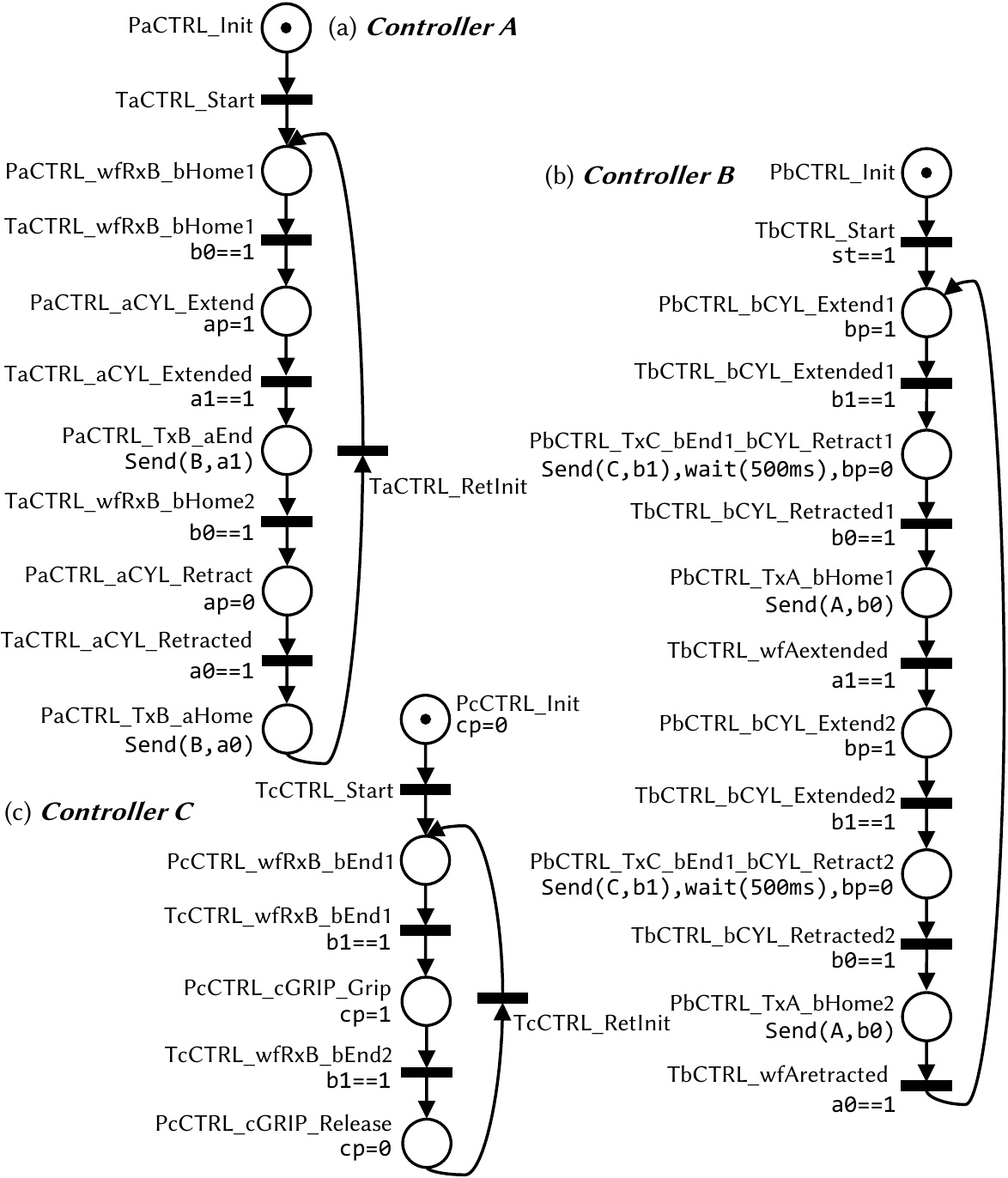}
	\caption{CIPN-based {distributed} controller of a 2-DOF pneumatic manipulator.}
	\label{fig:CIPNdistrCtrl}
\end{figure}

CIPN-based models of three LCs are shown in Fig.~\ref{fig:CIPNdistrCtrl}. Initially, cylinders $A$ and $B$ are fully retracted, and gripper $C$ released---in this state the manipulator is ready to begin its work cycle. The initial work cycle of the manipulator is started by pressing the start switch (\verb!st==1!), after which operation is fully automated. First, cylinder $B$ extends towards the workpiece picking position (due to actuation command \verb!bp=1!). Once cylinder $B$ reaches its end position (\verb!b1==1!), gripper $C$ is commanded gripping (\verb!cp=1!). Controller $B$ waits for $500~ms$ for the part to be gripped.\footnote{The gripper $C$ does not have end position sensing due to size constraints; thus a timed delay is used to permit secure gripping/releasing of the workpiece.}
Then, cylinder $B$ retracts (due to command \verb!bp=0!), and once it reaches home position (\verb!b0==1!), cylinder $A$ extends (due to command \verb!ap=1!). After reaching its end position (\verb!a1==1!), cylinder $B$ extends towards the placing position (due to command \verb!bp=1!). Once it reaches its end position (\verb!b1==1!), gripper $C$ is commanded release of the workpiece (command \verb!cp=0!). $500~ms$ later, cylinder B retracts (\verb!bp=0! followed by \verb!b0==1!), after which cylinder $A$ retracts (\verb!ap=0! followed by \verb!a0==1!). The manipulator returnees into its initial state, after which the next cycle is automatically executed. Signals (i.e., sensors outputs and actuator inputs) are allocated to LCs according to their physical proximity: $\{\verb!ap!,\verb!a0!,\verb!a1!\}$ are mapped to controller A (i.e., $LC_A$), $\{\verb!bp!,\verb!b0!,\verb!b1!,\verb!st!\}$ to controller B ($LC_B$), and $\{\verb!cp!\}$ to controller C ($LC_C$).

TPN models are obtained from these specifications as described in Section~\ref{sec:modeling}, but are omitted here due to their size. On the other hand, pneumatic cylinders are modeled as two-state plants with bounded, non-deterministic extending/retracting times obtained from experimental measurements. We extract timing parameters (i.e., bounds on time-to-transmit, time-to-acknowledge, and back-off timing) from experimental measurements---histograms for $10,000$ messages are shown in Fig.~\ref{fig:histograms}, for the employed low-cost ARM Cortex-M3-based controllers equipped with an IEEE~802.15.4-compliant transceiver. On the other hand, we obtain transceiver-related timings (e.g., back-off time during clear channel assessment) from the radio specifications~\cite{mrf}. While we verified a large number of safety and liveness properties for this setup, we~illustrate~verification and security patching on a more complex 3-DOF setup.

\begin{figure}[t]
	\includegraphics[width=0.49\textwidth,left]{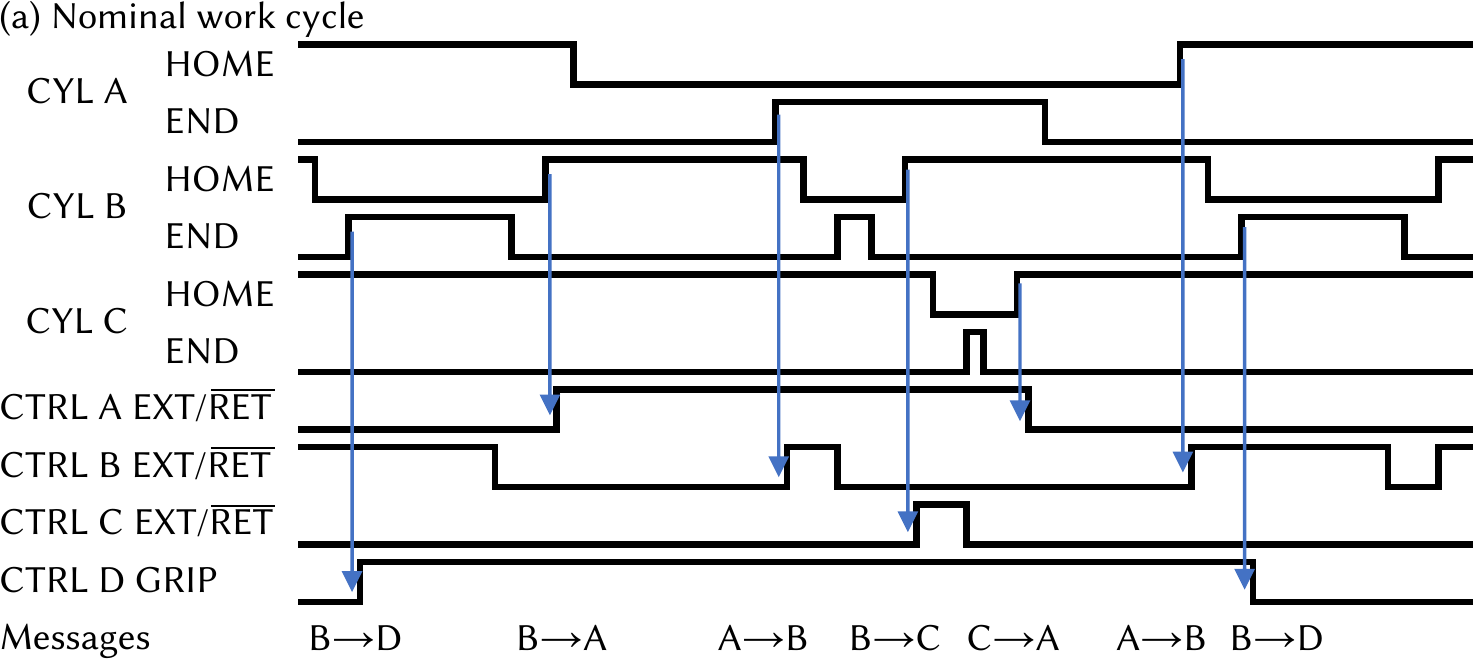}\vspace{5pt}
	\includegraphics[width=0.49\textwidth,left]{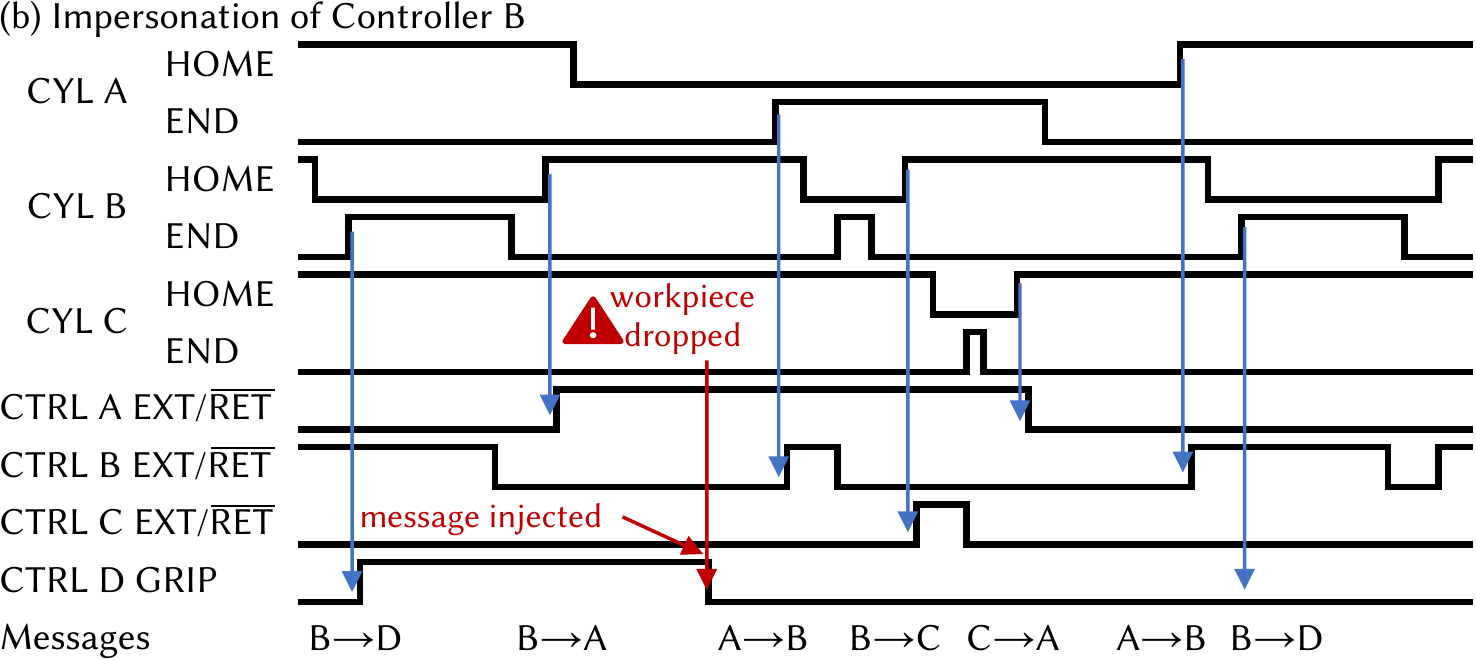}\vspace{5pt}
	\includegraphics[width=0.49\textwidth,left]{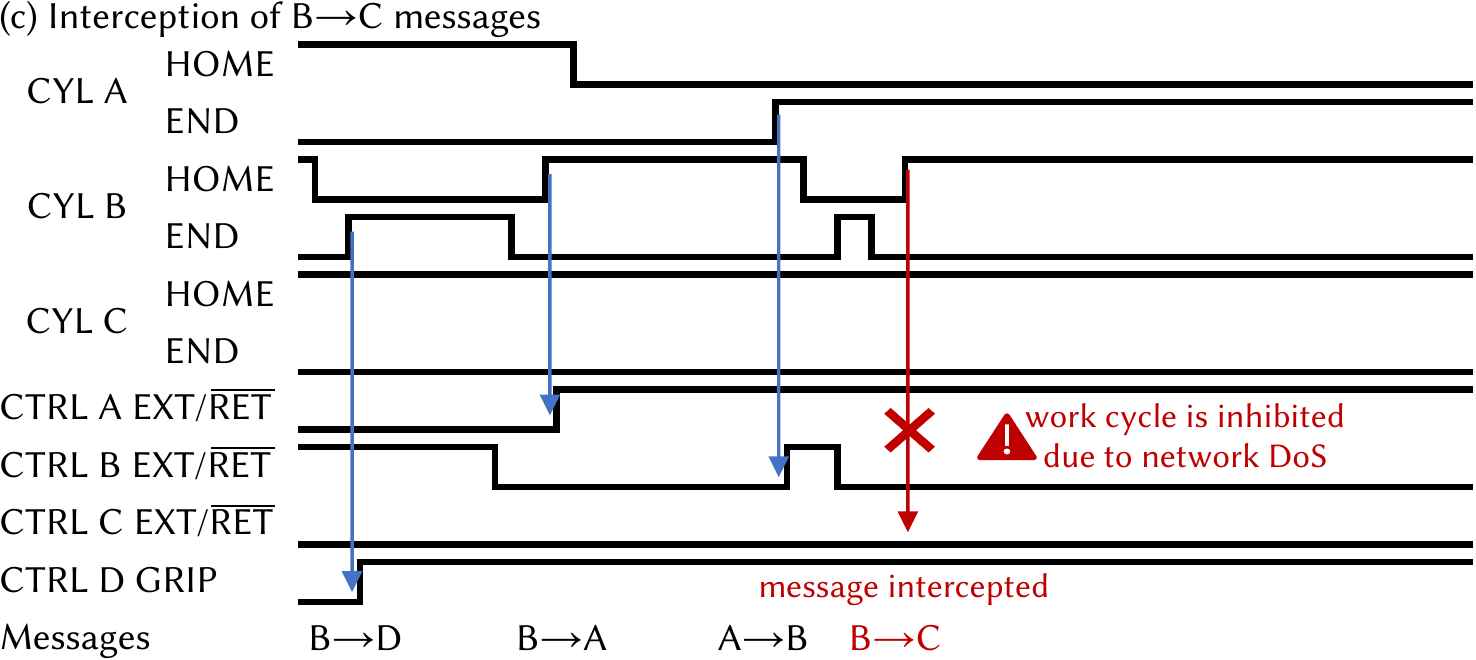}
	\caption{Sensing/actuation signal timings for a nominal pick~\&~place run (a), a run where a signal injection is performed resulting in a dropped workpiece (b), and a run where progress is inhibited due to a DoS attack (c). Messages exchanged by LCs are marked with blue arrows. $X$ axis is unlabeled as the speed of the workcycle can be controlled by regulating air pressure in the system and is thus not crucial.}
	\label{fig:plots}
\end{figure}

\subsection{3-DOF Industrial Pneumatic Manipulator}
A 3-DOF configuration of the described manipulator is shown in Fig.~\ref{fig:caseStudies}(b). The additional rotational DOF, provided by cylinder C, introduces an additional LC and increases the complexity of the LC coordination. This configuration may be used to prepare workpieces for painting by immersing them into a pool with cleaning solution, and returning them to the pick-up position for further processing by another machine.

Fig.~\ref{fig:caseStudies}(c,1) shows the physical setup for  this configuration; the upper portion of the manipulator is shown such that cylinders are visible. Fig.~\ref{fig:caseStudies}(c,2) shows the low-cost ARM Cortex-M3-based LC with the corresponding IEEE~802.15.4 transceiver. While the models are more complex than in the 2-DOF case, they are semantically similar and thus omitted. Fig.~\ref{fig:plots}(a) shows event timing---i.e., states of all sensing and actuation signals, for one sample pick-immerse-shake-return run; messages exchanged by LCs are denoted with blue arrows originating at the source event and terminating at the triggered event. Among the many safety liveness and QoC properties, we illustrate verification on the following examples.
\begin{property}\label{prop:caseStudyP1}
Gripper D is always gripped before cylinder B picks the workpiece; formally captured as, \verb!AG(M(PdGRIP_Gripped)==1 and!\\\verb!M(PbCTRL_bCYL_Retract1)==1)!.
\end{property}
\begin{property}\label{prop:caseStudyP2}
A workpiece is eventually processed, once the work cycle is started. Formally, \verb!M(PbCTRL_bCyl_Extend1)==1-->!\\\verb!(M(PcCTRL_cGRIP_Release)==1!.
\end{property}

When no security mechanisms are employed, we verified violation of these properties.  Property~\ref{prop:caseStudyP1} is violated due to a possible impersonation attack at the gripper controller; an attacker may send the command to release the workpiece before it was returned to the return position. Fig.~\ref{fig:plots}(b) shows signal timings acquired on a sample cycle run in which the workpiece is dropped due to a maliciously injected command to release the gripper (potentially causing mechanical damage to the workpiece and/or the manipulator).

However, if transmissions are authenticated, and the model adjusted correspondingly as described in Section~\ref{subsubsec:authentication}, this vulnerability is alleviated. We applied a software security patch by including the \emph{mbed TLS} (Secure Sockets Layer) library that our IIoT controllers are fully compatible with. Signing a $128~bit$ message authentication code over one transmitted signal incurs computational overhead of $\sim100~\mu s$ on the employed low-cost ARM Cortex-M3-based LCs; this practically negligibly slows down manipulator's work cycle, while providing security guarantees. Hence, Property~\ref{prop:caseStudyP1} is satisfied following this security patch.

Property~\ref{prop:caseStudyP2} is violated due to the possibility of a DoS attack that infinitely delays progress. From the model's perspective, this attack does not cause a deadlock---while the \emph{physical} process is stalled, the \emph{cyber} process is in fact livelocked reattempting to access the channel (i.e., same places are revisited and same transitions fire infinitely often). Fig.~\ref{fig:plots}(c) shows signal timings acquired on a sample run where a DoS attack is launched by enabling carrier transmission on the attacker's transceiver, in order to jam messages after the workpiece was picked up from the immersion pool. As described in Section~\ref{subsubsec:DoSdetection}, wireless control nodes can keep track of unsuccessful medium access attempts, and promptly halt operation when a DoS attack is detected.
%
In such cases, distributing the information about DoS detection requires a secure channel, which we do not consider in this work.



%% file: Conclusion.tex


\section{Conclusion}
\label{sec:conclusion}
In this paper, we have developed a framework for security analysis of distributed sequential control systems captured as widely-adopted CIPN-based models. As CIPNs do not support verification of formal safety properties in the presence of attacks, we transform controller models into TPNs that inherently enable this verification by supporting non-deterministic timed transition as well as non-deterministic choice among transitions; this is crucial as it imposes minimal assumptions on adversarial actions. We have shown how a model of a network-based attacker can be integrated into the non-deterministic communication channel model, and have verified violation of safety properties in presence of attacks. 

Additionally, we have shown how results of verification can be used to pinpoint vulnerabilities in the control software implementation, and suggest security patches to alleviate impact of these vulnerabilities on control performance; we have also provided the loop back to the modeling stage enabling verification of the same safety properties that are now satisfied due to the use of appropriate security mechanisms. Finally, we have evaluated our framework on an industrial case study of a realistic scale and complexity.